\documentclass[aps,pra,floatfix,twocolumn,amsmath,amssymb,showpacs,10pt]{revtex4-1}

\usepackage{graphicx}
\usepackage{pst-all,pstricks-add}

\newcommand{\ket}[1]{\lvert #1 \rangle}           
\newcommand{\bra}[1]{\langle #1 \lvert}           
\newcommand{\expv}[1]{\langle #1 \rangle}           

\newcommand{\rhoop}{\hat{\rho}}

\newcommand{\rhoi}{\hat{\rho}_\mathrm{i}}
\newcommand{\rhoix}{\hat{\rho}_\textrm{i $(x)$}}
\newcommand{\rhoo}{\hat{\rho}_\mathrm{0}}
\newcommand{\rhof}{\hat{\rho}_\mathrm{f}}

\newcommand{\rhofs}{\hat{\rho}_\textrm{f s}}
\newcommand{\rhoprep}{\hat{\rho}_\textrm{prep}}
\newcommand{\rhox}{\hat{\rho}_\textrm{$(x)$}}
\newcommand{\rhoxprime}{\hat{\rho}_\mathrm{(x^\prime)}}
\newcommand{\rhoNx}{\hat{\rho}_\textrm{$(N-x)$}}
\newcommand{\rhofx}{\hat{\rho}_\textrm{f $(x)$}}
\newcommand{\rhofsderiv}{\hat{\dot{\rho}}_\textrm{f s}}
\newcommand{\Hs}{H_\textrm{s}(\lambda)}
\newcommand{\Hsopt}{H_\textrm{s opt}(\lambda)}
\newcommand{\Hsoptpure}{H_\textrm{s pure opt}(\lambda)}
\newcommand{\Hoptpure}{H_\textrm{pure opt}(\lambda)}

\newcommand{\Hoptind}{H_\textrm{opt ind}(\lambda)}
\newcommand{\Hx}{H_\mathrm{(x)}}
\newcommand{\Aop}{\hat{A}}
\newcommand{\score}{\hat{L}}
\newcommand{\scorex}{\hat{L}_{(x)}}

\newcommand{\sigmax}{\hat{\sigma}_x}
\newcommand{\sigmay}{\hat{\sigma}_y}
\newcommand{\sigmaz}{\hat{\sigma}_z}
\newcommand{\sigman}{\hat{\sigma}_n}
\newcommand{\lambdaest}{\lambda_\mathrm{est}}
\newcommand{\estimate}{\tilde{\lambda}}
\newcommand{\myvector}[1]{\boldsymbol{\mathrm{#1}}}
\newcommand{\unitvec}[1]{\boldsymbol{\hat{#1}}}
\newcommand{\uprep}{\hat{U}_\mathrm{prep}}
 
\DeclareMathOperator{\Trace}{Tr}
\DeclareMathOperator{\variance}{var}
\DeclareMathOperator{\mse}{mse}

\begin{document}

\author{David Collins}
\affiliation{Department of Physical and Environmental Sciences, Colorado Mesa University, Grand Junction, CO 81501}
\email{dacollin@coloradomesa.edu}

\title{Mixed state Pauli channel parameter estimation}

\begin{abstract}
 The accuracy of any physical scheme used to estimate the parameter describing the strength of a single qubit Pauli channel can be quantified using standard techniques from quantum estimation theory.  It is known that the optimal estimation scheme, with $m$ channel invocations, uses initial states for the systems which are pure and unentangled and provides an uncertainty of ${\cal O}(1/\sqrt{m})$. This protocol is analogous to a classical repetition and averaging scheme. We consider estimation schemes where the initial states available are not pure and compare a protocol involving quantum correlated states to independent state protocols analogous to classical repetition schemes. We show, that unlike the pure state case, the quantum correlated state protocol can yield greater estimation accuracy than any independent state protocol. We show that these gains persist even when the system states are separable and, in some cases, when quantum discord is absent after channel invocation. We describe the relevance of these protocols to nuclear magnetic resonance measurements. 
\end{abstract}

\pacs{03.65.Ta, 03.67.-a,03.65.Ud}

\maketitle


\section{Introduction}
\label{sec:intro}

Quantum estimation considers situations in which a quantum system or evolution operation depends on one or more parameters, the values of which are to be estimated as accurately as possible. Such estimation will have to be accomplished by subjecting physical systems to physical procedures and consequently the accuracy of estimates will be governed by the relevant laws of physics. This realization has resulted in substantial study of constraints and quantum advantages available for estimation in the context of quantum theory and has yielded results with important implications for quantum metrology~\cite{hellstrom76,caves80,shapiro91,caves93,braunstein94,giovannetti04,sarovar06,giovannetti06,paris09,berry09,oloan10,escher11,giovannetti11}.

In general, any quantum estimation procedure considers a quantum operation of known form but which depends on one or more parameters.  The focus of this article is the Pauli operation (or channel), $\hat{\Gamma}(\lambda)$, which acts on a single qubit. This transforms a system, initially in the state $\rhoop$, via
\begin{equation}
	\rhoop \stackrel{\hat{\Gamma}(\lambda)}{\mapsto}  \rhof(\lambda) := \left(1-\lambda \right) \rhoop + \lambda \sigman \rhoop \sigman
	\label{eq:phaseflip}
\end{equation}
where $0 \leqslant \lambda \leqslant 1$ and $\sigman$ is a Pauli operator with $n$ represent a direction in space. For example, $n=x$ represents the bit-flip operation and $n=y$ is the phase-flip operation. The assumption is that the direction $n$ is known but the parameter $\lambda$ is unknown. The task is to estimate $\lambda$ with minimal uses of physical resources, such as qubits, number of applications of the channel or ancillary quantum operations. 

Channels of this type are important for the following reasons. First, the three channels for which $n=x,y$ and $z$ are sufficient to generate the types of errors that are considered in quantum information processing~\cite{nielsen00}. Estimating their strengths via the parameters will be relevant for practical quantum information processing protocols.

Second, specific examples of such channels appear in contexts such as nuclear magnetic resonance (NMR), in which the phase-flip operation generates the dephasing (or $T_2$) processes~\cite{slichter96,jones11}. Here, the density operator of a single spin-1/2 system evolves with time $t$ as
\begin{equation}
	\rhoop(t) = \begin{pmatrix}
	                \rho_{00} & \rho_{01}e^{-t/T_2} \\
									 \rho_{10}e^{-t/T_2}\;  & \rho_{11}
	              \end{pmatrix}
\end{equation}
where $T_2$ is a constant with units of time. This is related to the phase-flip parameter by 
\begin{equation}
	\lambda = (1- e^{-t/T_2})/2.
  \label{eq:lambdaT2}
\end{equation}
Measuring or estimating the phase-flip parameter is thus equivalent to measuring $T_2.$ 

Third, quantum parameter estimation can illustrate fundamental differences between classical and quantum worlds. For example, consider two methods of estimating the parameter governing unitary phase shifts~\cite{giovannetti06}. Applying the unitary phase operation once to each of $n$ independent quantum systems, yields an optimal uncertainty of ${\cal O}(1/\sqrt{n})$, also called the standard quantum limit. This is analogous to a classical repetition and averaging procedure. A method which applies the unitary phase operation once to each of $n$ quantum systems which are initially in a particular entangled state yields an uncertainty of ${\cal O}(1/n)$.  This is the Heisenberg quantum limit and is optimal for this type of operation~\cite{giovannetti06, zwierz10, zwierz12}. Evidently the resource of entanglement, occurring in quantum systems, has yielded an advantage here.

Parameter estimation for Pauli channels has been considered in the context of finding an optimal scheme using all possible input states of the quantum systems involved~\cite{fujiwara03,sarovar06,oloan07,ji08,chiuri11,ruppert12}. The results, reviewed in section~\ref{sec:input}, are that the optimal accuracy can be attained when the channel is invoked multiple times on independent and unentangled systems, each of which is initially in a pure state~\cite{fujiwara03}. The manifestly quantum resource of entanglement has no role to play in this circumstance.  However, in situations such as solution state NMR, initial states are highly mixed. The question remains of whether an independent channel invocation approach is still optimal when the only initial states available are mixed. This variant of quantum parameter estimation has been investigated for unitary operations~\cite{dariano05,boixo08b,modi11,datta11} where advantages of the type using entangled pure states survived with mixed, correlated initial states. 

In this article we consider the question of Pauli channel parameter estimation when the initial states of the available quantum systems are mixed. We aim to compare the accuracy of an independent channel use approach to one where correlated or entangled states are used. This will be important for Pauli channels in solution state NMR and other situations where the initial states are inherently mixed.  Particularly, for NMR it is not practical to purify the state by a process of measurement followed by selection based on measurement outcomes; the full range of available initial does not include pure states. Our approach is similar to that of Modi, et.~al.~\cite{modi11} but the results are subtly different, particularly in terms of the relationship between the presence of entanglement or discord when improved accuracy is attained. 

We shall focus on the phase-flip channel but the results hold for any Pauli channel. Section~\ref{sec:genestimation} reviews the general framework for quantum parameter estimation and general bounds for phase-flip parameter estimation. Section~\ref{sec:input} describes initial states available and gives the optimal independent channel use estimation accuracy; this provides a ``classical'' benchmark against which other schemes can be compared. Section~\ref{sec:correst} outlines a particular approach that uses correlated quantum states and gives central new results which form the foundation our the rest of the article. Sections~\ref{sec:correstone} and~\ref{sec:correstmany} describe two cases of the main result with notably different outcomes. Finally section~\ref{sec:entangdisc} discusses the presence of correlations such as entanglement or quantum discord in our estimation schemes.



\section{Quantum parameter estimation}
\label{sec:genestimation}

\begin{figure}[h]
  \includegraphics[scale=.625]{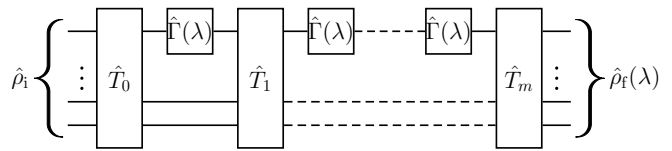} 
  \caption{Schematic of a quantum estimation procedure using $n$ quantum systems, each represented by a horizontal line. This procedure uses $m$ invocations of the parameter-dependent operation $\hat{\Gamma}(\lambda),$ interspersed with parameter-independent unitary transformations $\hat{T}_1, \ldots,  \hat{T}_{m}.$ The initial state of all quantum systems is $\rhoi$ and the pre-measurement state  $\rhof(\lambda).$ Measurements on individual systems follow the final unitary $\hat{T}_m.$
           \label{fig:generalscheme}}
 \end{figure}

Typically a physical quantum estimation procedure commences with various quantum systems prepared in known initial states. The parameter dependent operation is applied multiple times to some  of these systems, possibly interspersed with ancillary parameter-independent quantum operations. The physical procedure terminates with measurements on the individual quantum systems. The parameter is inferred by data processing from the measurement outcomes. 
 Fig.~\ref{fig:generalscheme} illustrates a generic quantum estimation scheme which uses $m$ invocations of the operation and $n$ quantum systems.
The general probabilistic nature of the outcomes of measurements on quantum systems means that the parameter inferences will fluctuate about the true parameter value. A key goal of any parameter estimation scheme is to minimize these fluctuations. 

Various estimation schemes  use differing resources (e.g.\ number of qubits, number of channel invocations, number of two qubit entangling unitaries) and a fair comparison requires precise specification of the resource costs. A common assessment of the resource cost uses a black-box query count within a quantum circuit~\cite{giovannetti06,vandam07,zwierz10,zwierz12} such as illustrated in Fig.~\ref{fig:generalscheme}. In this approach, used in this article, the only relevant resource cost is the number of times, $m,$ that the phase-flip operation appears in the quantum circuit for the protocol. We will only compare schemes that use the same number of channel invocations. 

In the scheme of Fig.~\ref{fig:generalscheme}, $\rhoi$ can be assumed to be a product of states for the individual qubits since an potential entangled state can be generated from this via an appropriate unitary, which can be subsumed into $\hat{T}_0.$ Given this, estimation schemes can broadly be grouped into two classes. 

In the first, called an \emph{independent channel use protocol,} the channel is applied once to each of $m$ qubits, which are initially in a product state. This is followed by measurements. In the context of Fig.~\ref{fig:generalscheme} this means that all of $\{ \hat{T}_1, \ldots,  \hat{T}_{m} \}$ are non-entangling amongst the qubits (a non-entangling unitary transforms any state which is a product of states for individual quantum systems to another such product state). In this protocol $\hat{T}_0$ is a product of single qubit unitaries while $\hat{T}_1, \ldots, \hat{T}_m$ are products of single qubit unitaries sandwiched between two qubit swap gates. Throughout the procedure of this protocol the state of the entire system is always a product of states for the individual qubits and there are no quantum correlations present. The independent channel use protocol is analogous to the classical strategy of invoking the physical operation independently on multiple physical systems; this generally enhances estimation accuracy. 

In the second approach, called a \emph{correlated state protocol} and only available to quantum systems, at least one of $\{ \hat{T}_1, \ldots, \hat{T}_m\}$ is an entangling unitary and, at some point during the procedure, the system will not be in a product state. 

The task of schemes falling into either class is to find admissible unitaries $\{ \hat{T}_1, \ldots, \hat{T}_m\}$ that maximize the estimation for fixed resource costs.  The question is whether there is some correlated state protocol that provides better estimation accuracy that any independent state protocol. If so, it would indicate that some quantum resources, possibly generated by an entangling unitary, have been harnessed so as to outperform any estimation protocol merely analogous to the classical strategy of repetition, as employed by an independent channel use protocol.

\subsection{Quantum Fisher information}

Quantifying the accuracy of quantum estimation procedures is done using the following general scheme. Denoting the outcome of the measurement on the $j^\mathrm{th}$ system by $x_j,$ the collective outcomes of the physical procedure of Fig.~\ref{fig:generalscheme} can be represented by $\{ x_1, \ldots, x_n\}$. The parameter is estimated by applying these an \emph{estimator function}, $\lambdaest,$ resulting in the estimate, $ \tilde{\lambda}= \lambdaest(x_1, \ldots, x_n)$. As a result of statistical fluctuations in measurement outcomes, inherent in quantum systems, the same estimator will yield possibly different results from one run to another which uses an identical sequence of quantum operations. The statistical accuracy of the estimate will therefore  rely on the choice of estimator together with the probability distribution for the measurement outcomes.  One means of quantifying the accuracy of this estimate uses the mean square error,
\begin{equation}
	\mse{(\estimate)}:= \left< \left( \estimate - \lambda \right)^2\right>
\end{equation}
where the angle brackets indicate the mean over all possible measurement outcomes. The possible estimators consider in this article will be \emph{unbiased estimators}, i.e.\ $\expv{\estimate} = \lambda$, and here $\mse{(\estimate)} = \variance{(\estimate)},$ where the variance in the estimate is
\begin{equation}
	\variance{(\estimate)}:= \left< \left( \estimate - \expv{\estimate} \right)^2\right>.
\end{equation}
The variance is the fundamental entity which quantifies the accuracy of the estimate. A fundamental result~\cite{cramer46,paris09,oloan10} is that, \emph{regardless of the estimator function, the variance is bounded from below via the Cram\'{e}r-Rao bound}
\begin{equation}
	\variance{(\estimate)} \geqslant \frac{1}{F(\lambda)}
  \label{eq:classicalcrb}
\end{equation}
where $F(\lambda)$ 
\begin{equation}
	F(\lambda):= \int 
	             \left[ 
	               \frac{\partial \ln{p(x_1, \ldots x_n|\lambda)}}{\partial \lambda}
	             \right]^2\;
	             \;
	             \mathrm{d}x_1 \ldots \mathrm{d}x_n.
	\label{eq:classicalfisher}
\end{equation}
is the Fisher information associated with the probability distribution (conditional upon $\lambda$) for the measurement outcomes, $p(x_1, \ldots x_n|\lambda)$. If each measurement outcome is one of a set of discrete possibilities, the integral is replaced by a summation.

It can be shown that under general conditions, there is always an estimator~\cite{cramer46}, which asymptotically saturates the bound of Eq.~\eqref{eq:classicalcrb}. Thus in classical parameter estimation the Fisher information quantifies accuracy of the estimate and when comparing estimation schemes, that with the greater Fisher information is more successful than any competitor with a lower Fisher information. 

Given a particular estimation procedure with Fisher information, $F(\lambda),$ an obvious and conventional classical strategy for improvement is to repeat the identical procedure multiple times independently. Mathematically this is equivalent to sampling the same probability distribution in an independent and identical fashion.  If this is done $m$ times then the Fisher information is $mF(\lambda)$; this governs the typical behavior as a function of $m$ in classical parameter estimation~\cite{oloan10}. For example, in the context of parameter estimation in classical physical systems, the same physical procedure can be applied to $m$ identical systems. The resulting mean square error is reduced by a factor of $m.$

For parameters pertaining to the evolution of quantum systems, the probability of measurement outcomes is determined by: (i) the choice of measurement and (ii) the choice of initial state $\rhoi$, which determines the final state $\rhof(\lambda).$ Therefore, the classical Fisher information for any given final state depends on the choice of the measurement. However, considering all possible quantum measurement procedures on the system and any possible ancillary quantum systems shows that the classical Fisher information satisfies the \emph{quantum Cram\'{e}r-Rao bound,}
\begin{equation}
	F(\lambda) \leqslant H(\lambda)
	\label{eq:quantumcrb}
\end{equation}
where the quantum Fisher information $H(\lambda)$ is \emph{independent of the choice of measurement}~\cite{braunstein94,paris09,oloan10}. The quantum Fisher information depends only on $\rhof(\lambda)$ and is calculated via~\cite{paris09}
\begin{equation}
	H(\lambda) = \Trace{\left[ \rhof(\lambda) \score^2(\lambda)\right]}
	\label{eq:quantumfisher}
\end{equation}
where the symmetric logarithmic derivative (SLD) or score operator, $\score(\lambda)$ is defined implicitly via
\begin{equation}
	\frac{\partial \rhof(\lambda)}{\partial \lambda} = \frac{1}{2}\;
	                                                   \left[
	                                                     \score(\lambda)
	                                                     \rhof(\lambda)
	                                                     +
	                                                     \rhof(\lambda)
	                                                     \score(\lambda)
	                                                   \right].
  \label{eq:slddefintion}
\end{equation}
A useful alternative to Eq.~\eqref{eq:quantumfisher} is
\begin{eqnarray}
 H(\lambda) & = & \Trace{\left[ \rhof(\lambda) \score^2(\lambda)\right]} \nonumber \\
            & = & \frac{1}{2}\; \Trace{\left[ \score(\lambda) \rhof(\lambda) \score(\lambda)
                                       + \rhof(\lambda) \score^2(\lambda) \right]} \nonumber \\
            & = & \Trace{\left[  \frac{\partial \rhof(\lambda)}{\partial \lambda} \score(\lambda) \right]}.
            \label{eq:quantumfishertwo}
\end{eqnarray}
There exist methods~\cite{paris09} for computing the SLD based on eigenvalue and eigenstate decompositions of $\rhof(\lambda)$ and conditions for the existence of measurements which saturate the bound of Eq.~\eqref{eq:quantumcrb}. In general there does exist a measurement procedure which asymptotically attains this bound~\cite{barndorff00}.  Quantum Fisher information then serves as a measure of the possible accuracy of the estimation procedure.  The actual measurement and data processing of the outcomes are irrelevant here and all that matters is the final state of the system, $\rhof(\lambda).$ The central task in any quantum parameter estimation scheme is to determine the initial state for the quantum system which maximizes the resulting quantum Fisher information and, if possible, the associated measurement which yields a classical Fisher information equal to the maximum quantum Fisher information.  This reduces the task of designing quantum parameter estimation schemes to that of tailoring $\rhof(\lambda)$ so as to maximize the quantum Fisher information.

\subsection{Determining the SLD}

Computing the SLD can be done by computing eigenstates and eigenvalues of $\rhof(\lambda)$  but this generally difficult. However, for certain operators, there are methods for computing the SLD which do not require the eigenvalues and eigenstates. 

\textit{Proposition 1:} Let $\hat{A}$ be a diagonalizable operator on a two dimensional vector space such that $\Trace{\Aop} \neq 0.$ Let $\alpha:= \Trace{(\Aop^2)} - (\Trace{\Aop})^2.$ Then $\frac{\partial\Aop}{\partial \lambda} = (\score \Aop +\Aop \score)/2$ where 
\begin{equation}
	\hat{L} = \frac{1}{\Trace{\Aop}}\; 
	          \left[ 2\; \frac{\partial\Aop}{\partial \lambda} 
	                - \frac{\partial \ln{(\Trace{\Aop})}}{\partial \lambda}\; \Aop 
					  \right]
	\label{eq:sldtwobytwofirst}
\end{equation}
if $\alpha =0$ and 
\begin{eqnarray}
	\hat{L} = & \dfrac{1}{\Trace{\Aop}}\; 
	            \left[ 2\; \dfrac{\partial\Aop}{\partial \lambda} 
	                - \dfrac{\partial \ln{\alpha}}{\partial \lambda}\; \Aop 
					    \right] \nonumber \\
						& + \dfrac{\partial}{\partial \lambda}\;
						    \left[
								  \ln{\alpha} - \ln{(\Trace{\Aop})}
								\right]
								\hat{I}
	\label{eq:sldtwobytwosecond}
\end{eqnarray}
if $\alpha \neq 0.$

\textit{Proof:} Denote the eigenvalues of $\Aop$ by $a_1$ and $a_2.$ The characteristic equation for $\Aop$ is $(\Aop - a_1 \hat{I})(\Aop - a_2 \hat{I}) = 0.$ Expanding this gives
\[ \Aop^2 - \Trace{[\Aop]} \Aop + a_1 a_2 \hat{I} =0.\]
Further algebra yields
\begin{equation}
 \Aop = \frac{1}{\Trace{\Aop}}\;
        \left( 
				     \Aop^2 
						 - \frac{1}{2}\;
						   \alpha
							 \hat{I}
				\right)
 \label{eq:proponeone}
\end{equation}
where we have used the fact that $\Trace{\Aop} \neq 0.$ Thus
\begin{align}
	\frac{\partial\Aop}{\partial \lambda} = & -\frac{1}{\Trace{\Aop}}\; 
	                                            \frac{\partial \ln{(\Trace{\Aop})}}{\partial \lambda}\;
																						  \left( \Aop^2 - \frac{\alpha}{2}\; \hat{I}
																							\right) \nonumber \\
																					 & + \frac{1}{\Trace{\Aop}}\; 
																						   \left(
																							  \frac{\partial\Aop}{\partial \lambda}\; \Aop
																								+ \Aop \frac{\partial\Aop}{\partial \lambda} 
																								- \frac{1}{2}\; 
																								  \frac{\partial\alpha}{\partial \lambda}\; \hat{I}
																							 \right).
 \label{eq:proponetwo}
\end{align}
If $\alpha =0$, Eq.~\eqref{eq:proponetwo} reduces to 
\[ \dfrac{\partial\Aop}{\partial \lambda} =  -\dfrac{1}{\Trace{\Aop}}\; 
	                                            \dfrac{\partial \ln{(\Trace{\Aop})}}{\partial \lambda}\;
																						  \Aop^2 
																						+ \dfrac{1}{\Trace{\Aop}}\; 
																						   \left(
																							  \dfrac{\partial\Aop}{\partial \lambda}\; \Aop
																								+ \Aop \dfrac{\partial\Aop}{\partial \lambda} 
																							 \right)
\] 
and by inspection, Eq.~\eqref{eq:sldtwobytwofirst} holds. On the other hand, if $\alpha \neq 0$ then Eq.~\eqref{eq:proponeone} can be inverted to give 
$\hat{I} = 2[\Aop^2 - \Trace{(\Aop}) \Aop]/\alpha$ and Eq.~\eqref{eq:proponetwo} becomes
\begin{align}
	\frac{\partial\Aop}{\partial \lambda} = & -\frac{\partial \ln{(\Trace{\Aop})}}{\partial \lambda}\;
																						  \Aop \nonumber \\
																					 & + \frac{1}{\Trace{\Aop}}\; 
																						   \left(
																							  \frac{\partial\Aop}{\partial \lambda}\; \Aop
																								+ \Aop \frac{\partial\Aop}{\partial \lambda} 
																								- \dfrac{1}{2}\; 
																								  \dfrac{\partial\alpha}{\partial \lambda}\; \hat{I}
																							 \right) \nonumber \\
																					 & - \frac{1}{\Trace{\Aop}}\;
																							  \frac{\partial \ln{\alpha}}{\partial \lambda}\; 
																						   \left(
																							  \Aop^2 - \Trace{(\Aop)}\; \Aop
																							 \right).
 \label{eq:proponethree}
\end{align}
Inspection of Eq~\eqref{eq:proponethree} shows that the SLD is given by Eq.~\eqref{eq:sldtwobytwosecond}. \hspace{\fill}$\Box$

Some situations require SLDs for operators that act on vector spaces of dimension greater than $2$. If such an operator can be decomposed into the sum of operators acting on orthogonal spaces then the quantum Fisher information can be determined in a piecewise fashion.

\textit{Proposition 2:}  Suppose that $\rhoop = \sum_k \rhoop_k$ where the supports of $\rhoop_j$ and $\rhoop_k$ are orthogonal whenever $j \neq k.$ Let $\score_k$ be the SLD for $\rhoop_k,$ i.e.\ $\frac{\partial \rhoop_k}{\partial \lambda} = (\score_k \rhoop_k + \rhoop_k \score_k)/2.$ Then the SLD for $\rhoop$ is $\score = \sum_{k} \score_k$ and the quantum Fisher information is $H = \sum_{k}H_k$ where $H_k = \Trace{(\frac{\partial \rhoop_k}{\partial \lambda} \score_k )}.$

\textit{Proof} Suppose that $\frac{\partial \rhoop_k}{\partial \lambda} = (\score_k \rhoop_k + \rhoop_k \score_k)/2.$ Then 
\begin{align}
	\frac{\partial \rhoop}{\partial \lambda} & = \sum_k \frac{\partial \rhoop_k}{\partial \lambda} \nonumber \\
	                                          & = \sum_k \frac{1}{2}\; 
																						    \left( 
																								 \score_k \rhoop_k + \rhoop_k \score_k
																								\right).
\end{align}
The support of $\score_k$ is either the same as or a subspace of the support of $\rhoop_k$. Thus if $j \neq k$ then $\score_j \rhoop_k =0$ and 
\begin{align}
	\frac{\partial \rhoop}{\partial \lambda} & = \frac{1}{2}\;
	                                           \left( 
																						   \sum_j \score_j \sum_k \rhoop_k
																							+  \sum_k \rhoop_k \sum_j \score_j
																						  \right)
																						 \nonumber \\
																					 & = \frac{1}{2}\;
	                                           \left(
																						  \score \rhoop + \rhoop \score
																						 \right).
\end{align}
The result regarding the quantum Fisher information follows by applying similar reasoning to Eq.~\eqref{eq:sldtwobytwosecond}. \hspace{\fill}$\Box$

\subsection{Bounds on the quantum Fisher information for phase-flip parameter estimation}
\label{sec:genqfi}

\begin{figure}[ht]
  \includegraphics[scale=.80]{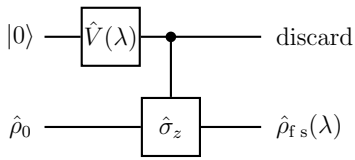} 
  \caption{Unitary representation of a phase-flip channel. The lower qubit is the channel qubit and the upper an ancilla qubit, which is required to be in the initial state, $\ket{0}$. After evolution under the two indicated unitary operations, the ancilla qubit is ignored or discarded. 
           \label{fig:extendedchannel}}
 \end{figure}

A useful tool for addressing bounds on the quantum Fisher information for the phase-flip operation, already considered elsewhere~\cite{fujiwara03,sarovar06,ji08}, involves a representation of the channel using an additional ancilla qubit as illustrated in Fig.~\ref{fig:extendedchannel}.

This pair of qubits is subjected to the indicated \emph{unitary} operations where  the additional ``\emph{coin-toss}'' unitary operation is
\begin{equation}
	\hat{V}(\lambda) : = \begin{pmatrix}
	                       \sqrt{\lambda} & -\sqrt{1-\lambda} \\
	                       \sqrt{1-\lambda} & \sqrt{\lambda}
	                     \end{pmatrix}.
\end{equation}
Following the controlled-$Z$ unitary, the ancilla qubit is discarded. Mathematically, this amounts to taking the partial trace over the state space for the ancilla qubit. It is straightforward to show that the resulting transformation of the channel qubit is identical to the phase-flip operation.

The quantum Fisher information of the phase-flip channel alone is bounded from above by the quantum Fisher information of the entire extended channel of Fig.~\ref{fig:extendedchannel} since the set of all possible measurements on both qubits includes the set of all measurements made on the channel qubit alone.   When evaluating the quantum Fisher information of the extended channel the  controlled-$Z$ unitary and the initial state of the channel qubit can be collectively absorbed into a general measurement procedure since they are parameter-independent. The only relevant entities are the coin-toss unitary and the initial state of the ancilla qubit.  Thus the quantum Fisher information for the extended channel is exactly that for the coin-toss unitary applied to the ancilla in the state $\ket{0}$. Proposition~1 and Eq.~\eqref{eq:quantumfishertwo} apply here and give that the quantum Fisher information for the extended channel is $1/\lambda(1-\lambda).$ 

Multiple uses of the phase-flip operation can be represented by a similar extension involving multiple coin-toss unitaries and ancilla qubits, each in the initial state $\ket{0}.$ According to the scheme above, these parts of the procedure occur independently of each other and it is straightforward to show that the quantum Fisher information of the extended channel (and thus the upper bound on any estimation procedure using any channel input states) using $m$ phase-flip operations is $m/\lambda(1-\lambda).$ This result is independent of the state of the channel qubits and also of any unitaries $\hat{T}_1, \ldots, \hat{T}_m$ that are applied to the channel qubits between channel invocations. Thus \emph{for any parameter estimation scheme involving $m$ channel invocations the quantum Fisher information satisfies}
\begin{equation}
	H(\lambda) \leqslant \frac{m}{\lambda(1-\lambda)}.
	\label{eq:absolutebound}
\end{equation}
This absolute upper bound has appeared elsewhere in a more general context~\cite{fujiwara03}.


\section{Input state choices and constraints}
\label{sec:input}

The quantum Fisher information partly depends on the choice of the initial state for the system. It will be essential to provide a general mathematical description of the various initial states; the quantum Fisher information and possible constraints on initial states will be expressed partly in terms of parameters appearing in this description. 

\subsection{General product input states}

In general the initial state of the system can be assumed to be a product of states for individual qubits; initial entangled states can then be accommodated via appropriate choice of $\hat{T}_0$. The general density operator for an individual qubit has form 
\begin{equation}
	\rhoop := \frac{1}{2}\; \left( 
	                        \hat{I} + \myvector{r} \cdot \hat{\myvector{\sigma}}
	                       \right)								
\end{equation}
where $\myvector{r} \cdot \hat{\myvector{\sigma}} = r_x \sigmax + r_y \sigmay + r_z \sigmaz$ and $\myvector{r} = \left(r_x,r_y,r_z \right)$ is a conventional three dimensional real vector whose norm, or \emph{polarization}, satisfies $r \equiv \left| \myvector{r} \right| := \sqrt{r_x^2 + r_y^2 + r_z^2} \leqslant 1.$  We assume that all qubits have the same initial polarization. The only remaining possible differences between initial states arises from the orientations of $\myvector{r}$ for each qubit. By including appropriate single qubit rotations in $\hat{T}_0$, any differences in these orientations can be removed. Thus we assume that $\myvector{r}$ is identical for each qubit and therefore $\rhoi = \rhoo \otimes \rhoo \otimes \cdots \otimes \rhoo$ ($n$ factors) where
\begin{equation}
	\rhoo := \frac{1}{2}\; \left( 
	                        \hat{I} + \myvector{r}\cdot \hat{\myvector{\sigma}}
	                       \right)
	\label{eq:rhooneinitial}			
\end{equation}
for a particular given $\myvector{r}.$ The freedom to include single qubit rotations in $\hat{T}_0$ thus implies that the only relevant input state parameter for the quantum Fisher information is the polarization, $r$.

\subsection{Pure input states}

A subset of possible input states are those which are pure, i.e.\ $r=1.$ Consider a single application of the channel to a qubit in the initial state $\ket{+\unitvec{y}} = \left( \ket{0} + i \ket{1} \right) /\sqrt{2}$ with all $\hat{T}_j = \hat{I}.$ Straightforward application of proposition~1 and Eq.~\eqref{eq:quantumfishertwo} yields a single use pure state quantum Fisher information 
\begin{equation}
	\Hsoptpure = \frac{1}
	             {\lambda \left( 1-\lambda \right)},
	\label{eq:Hsingleoptimalpure}
\end{equation}
identical to that of the bound of Eq~\eqref{eq:absolutebound} with $m=1$ and therefore optimal. In fact, the projective measurement in the basis $\left\{ \ket{+\unitvec{y}}, \ket{-\unitvec{y}} \right\},$ where $\ket{\pm \unitvec{y}} = \left( \ket{0} \pm i \ket{1} \right) /\sqrt{2},$ is readily shown to yield a classical Fisher information equal to $\Hsoptpure.$

It follows that if this is repeated independently with $m$ channel invocations, each applied to a single qubit as described above, the resulting classical Fisher information is $m$ times that for a single channel use. But this saturates the bound of Eq.~\eqref{eq:absolutebound} and thus the optimal pure state quantum Fisher information with $m$ channel invocations is
\begin{equation}
	\Hoptpure = \frac{m}
	             {\lambda \left( 1-\lambda \right)}.
	\label{eq:Hmanyoptimalpure}
\end{equation}
Clearly for pure input states the optimal quantum Fisher information can be attained using product states (this is an example of the independent channel use protocol) . \emph{For pure input states no correlated state protocol can yield a better estimate than the particular independent channel use protocol describe in this section.} Evidently entanglement or quantum correlations between qubits cannot yield any advantages for phase-flip parameter estimation with initial pure states.

\subsection{Independent Channel Use Protocol}
\label{sec:indphaseest}

In the more general  independent channel use protocol the initial states are not pure. Suppose that  the operation is applied once to each of $m$ qubits, each of which is in the identical initial state, given by Eq.~\eqref{eq:rhooneinitial} and which appear leftmost in the mathematical representation of the entire system state. The final state of the collection of qubits is 
$\rhof = \rhofs \otimes \cdots \otimes \rhofs \otimes \rhoo \otimes \cdots \otimes \rhoo$ 
($m$ factors of $\rhofs$) where the final state for the individual qubits to which the operation is applied is $\rhofs = \left(1-\lambda \right) \rhoo + \lambda \sigmaz \rhoo \sigmaz.$ Let $\hat{L}_\textrm{s}$ be the SLD corresponding to $\rhofs$ and let $\Hs$ be the resulting quantum Fisher information for this single qubit. It is straightforward to show that the SLD for all $n$ qubits is $\hat{L}_\textrm{s} \otimes \hat{I} \otimes \cdots \otimes \hat{I} + \hat{I} \otimes \hat{L}_\textrm{s} \otimes \hat{I} \otimes \cdots \otimes \hat{I} + \cdots$ with one term for each qubit to which the operation is applied. Then Eq.~\eqref{eq:quantumfishertwo} implies that the quantum Fisher information for the entire system is
\begin{equation}
 H(\lambda) = m\; \Hs.
 \label{eq:independentqfi}
\end{equation}

The quantum Fisher information for a single channel use on one qubit can be computed by first determining the SLD. The conditions of proposition 1 apply and, since $\rhofs$ is not a pure state, $\alpha  = \Trace{(\rhofs^2)} - (\Trace{\rhofs}^2)\neq 0$. Thus
\begin{equation}
\hat{L}_\textrm{s} = 
 2 \left\{ 
   \rhofsderiv 
 - \dfrac{\Trace{\left[ \rhofsderiv \rhofs \right]}}{1- \Trace{\left[ \rhofs^2 \right]}}\; 
   \left( \hat{I} - \rhofs \right) 
  \right\}.
\end{equation}
Substituting into Eq.~\eqref{eq:quantumfishertwo} gives
\begin{equation}
	\Hs =  	2 \Trace{\left[ \rhofsderiv^2 \right]} - 2 \dfrac{\left( \Trace{\left[ \rhofsderiv \rhofs\right]}\right)^2}
						                                                {1- \Trace{\left[ \rhofs^2 \right]}}. 
	\label{eq:Hsingle}
\end{equation}
Applying this to $\rhofs$ as produced by the phase-flip channel and using  $\rhoo$ given by Eq.~\eqref{eq:rhooneinitial} yields
\begin{equation}
	\Hs = \frac{ 4 \left( 1-r_z^2 \right) \left( r^2 - r_z^2\right)}
	             {\left( 1-2\lambda \right)^2 \left( 1 - r^2\right) + 4\lambda(1-\lambda)\left( 1-r_z^2 \right) }.
\end{equation}
Thus the optimal single qubit, single channel use quantum Fisher information is attained when $r_z=0$ and is
\begin{equation}
	\Hsopt = \frac{4 r^2}
	             {1- \left( 1-2\lambda \right)^2r^2}.
	\label{eq:Hsingleoptimal}
\end{equation}
Thus, \emph{for $m$ independent channel uses} the optimal quantum Fisher information is
\begin{equation}
	\Hoptind = \frac{4 r^2\; m}
	             {1- \left( 1-2\lambda \right)^2r^2}.
	\label{eq:Hmanyindepoptimal}							
\end{equation}

One physical implementation which yields a classical Fisher information equal to that of $\Hsopt$ commences with $\rhoo =  \left( \hat{I} + r \sigmay \right)/2$ and follows a single channel invocation with a projective measurement in the basis $\left\{ \ket{+\unitvec{y}}, \ket{-\unitvec{y}} \right\}$ where $\ket{\pm \unitvec{y}} = \left( \ket{0} \pm i \ket{1} \right) /\sqrt{2}.$ A straightforward derivation using the probability distribution for the measurement outcomes and Eq.~\eqref{eq:classicalfisher} results in $F(\lambda) = \Hsopt.$

Eq.~\eqref{eq:Hmanyindepoptimal} gives the optimal quantum Fisher information in the independent channel use protocol when all qubits have initial polarization $r.$ This provides the benchmark against which correlated state protocol estimation schemes will be compared.


\section{Correlated State Protocol Parameter Estimation}
\label{sec:correst}

The results of section~\ref{sec:genqfi} show that for pure input states, any correlated state protocol  procedure cannot improve on the independent channel use protocol quantum Fisher information of Eq.~\eqref{eq:Hmanyoptimalpure}. However, for mixed input states, the result of Eq.~\eqref{eq:Hmanyindepoptimal} does not saturate the absolute bound of Eq.~\eqref{eq:absolutebound}. Here it is conceivable that a correlated state protocol procedure could improve upon the quantum Fisher information of  Eq.~\eqref{eq:Hmanyindepoptimal} given the same number of channel invocations.

\begin{figure}[ht]
  \includegraphics[scale=.80]{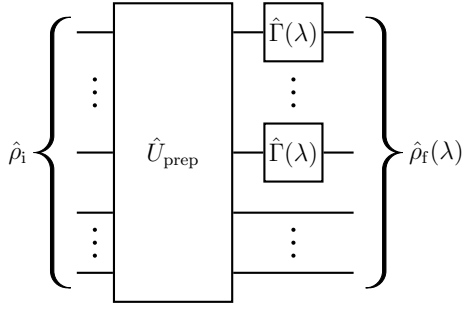} 
  \caption{Single preparation unitary scheme for improved parameter estimation. There are $n$ single qubits. A joint preparatory unitary, $\uprep$ precedes a phase-flip operation applied once to each of the uppermost $m$ qubits. This is of the form of the general scheme of Fig.~\ref{fig:generalscheme} where $\hat{T}_0 = \uprep, \hat{T}_m = \hat{I}$ and $\hat{T}_1, \ldots \hat{T}_{m-1}$ being swap gates between pairs of qubits. 
           \label{fig:preparedscheme}}
 \end{figure}

\begin{figure}[ht]
  \includegraphics[scale=.80]{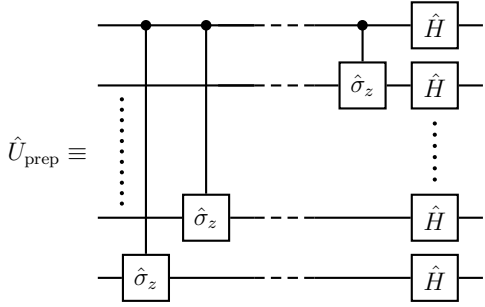} 
  \caption{The preparatory gate, $\uprep,$ consists of a collection of controlled-$Z$ operations applied to each distinct pair of qubits followed by a Hadamard transformation applied to each qubit. The preparatory gate is symmetric under interchange of any qubits since a controlled-$Z$ gate is symmetric between the pair of qubits on which it acts.
           \label{fig:uprep}}
 \end{figure}

Henceforth suppose that $r < 1$ and consider the correlated state protocol procedure as illustrated in Fig.~\ref{fig:preparedscheme}.  
Specifically suppose that  $\rhoo = (\hat{I}+ r \sigmay)/2$ and $\uprep$ is as illustrated in Fig.~\ref{fig:uprep}.
Denote the density operator for the system after the preparatory gate by $\rhoprep : =  \uprep \rhoi \uprep^\dagger.$ Then (see appendix~\ref{app:rhoprep}) 
\begin{equation}
  \rhoprep = \sum_{x=0}^{(N-1)/2} \rhox
\end{equation}
where $N= 2^n-1$ and 
%
\begin{equation}
	\begin{split}
	  \rhox = & 
		         \left[
		           \frac{f(x) + f(N-x)}{2}
              \right]\;
							\biggl[
		           \ket{x}\bra{x} + 
							 \ket{N-x}\bra{N-x}
              \biggr]
		        \\
	          & 
						+ i
						  \left[
		           \frac{f(x) - f(N-x)}{2}
              \right]\;
							\biggl[
		           \ket{x}\bra{N-x} -
							 \ket{N-x} \bra{x}
              \biggr].
	\end{split}
 \label{eq:rhoprepped}
\end{equation}
%
In Eq.~\eqref{eq:rhoprepped}, $\ket{x} = \ket{x_n\ldots x_1}$ where $x_n\ldots x_1$ is the binary representation of $x$, with the rightmost bit the least significant digit. Note that $\ket{N-x}$ can be obtained by flipping each bit value in $\ket{x}.$ Finally
\begin{equation}
	f(x) = \frac{(1+r)^j(1-r)^{n-j}}{2^n}
	\label{eq:fdefn}
\end{equation}
where $j$ is the number of zeroes in the bit string for $x$. The support of $\rhox$ is the two dimensional subspace spanned by $\left\{ 
\ket{x}, \ket{N-x} \right\}$ and, with one type of exception, if $x \neq x^\prime$ then $\rhox$ and $\rhoxprime$ have orthogonal supports. The only exceptions to this are that $\rhox = \rhoNx,$ a fact which is straightforward to show. Mathematically, the preparatory gate has engineered a splitting of the Hilbert space for the entire system into mutually orthogonal two-dimensional subspaces.

The preparation scheme is symmetric under interchange of qubits and without loss of generality, consider the situation where the phase-flip operation is applied once to each of the qubits corresponding to the $m$ least significant (or rightmost) digits in the computational basis representation. Then 
\begin{equation}
  \rhof = \sum_{x=0}^{(N-1)/2} \rhofx
\end{equation}
where $\rhofx$ results from applying the phase-flip operation to $\rhox.$ Here
\begin{equation}
	\begin{split}
	  \ket{x}\bra{x} & \mapsto \ket{x}\bra{x} \\
		\ket{N-x}\bra{N-x} 
		& \mapsto \ket{N-x}\bra{N-x}  \\
		\ket{x}\bra{N-x} 
		& \mapsto (1-2\lambda)^m\; \ket{x}\bra{N-x} \\
		\ket{N-x} \bra{x}
		& \mapsto (1-2\lambda)^m\;  \ket{N-x} \bra{x}		
	\end{split}
\end{equation}
and thus
%
\begin{equation}
	\begin{split}
	  \rhofx = & 
		         \left[
		           \frac{f(x) + f(N-x)}{2}
              \right]\;
							\biggl[
		           \ket{x}\bra{x} + 
							 \ket{N-x}\bra{N-x}
              \biggr]
		        \\
	          &
						 + i (1-2\lambda)^m\;
						  \left[
		           \frac{f(x) - f(N-x)}{2}
              \right]\;
						\\
						& \times
							\biggl[
		           \ket{x}\bra{N-x} -
							 \ket{N-x} \bra{x}
              \biggr].
	\end{split}
 \label{eq:rhofinalafterm}
\end{equation}
%
Each of $\rhofx$ has a two dimensional support and these are clearly orthogonal for distinct values of $x$ in the range $0,\ldots, (N-1)/2.$ Thus Proposition~2 gives that $H(\lambda) = \sum_{x=0}^{(N-1)/2} \Hx$ where $\Hx = \Trace{(\frac{ \partial \rhofx}{\partial \lambda} \scorex)}$
and $\scorex$ is the score operator for $\rhofx.$

Proposition~1 applies to $\rhofx$ since it has  two dimensional support and $\Trace{\rhofx} = f(x) + f(N-x) \neq 0.$ Direct calculation shows that $\Trace{(\rhofx)^2} - (\Trace{\rhofx})^2 = 0$ if and only if 
$\left[ f(x) - f(N-x) \right]^2\; 
 (1-2\lambda)^{2m} = 
 \left[ f(x) + f(N-x) \right]^2.$
The fact that $f(x) \geqslant 0$ implies that $\left[ f(x) + f(N-x) \right]^2 \geqslant \left[ f(x) - f(N-x) \right]^2.$ Furthermore $(1-2\lambda)^{2m} \leqslant 1.$ This implies $\Trace{(\rhofx)^2} - (\Trace{\rhofx})^2 = 0$ if and only if $(1-2\lambda) = \pm 1$ (i.e.\ $\lambda =0,1$) and $\left[ f(x) + f(N-x) \right]^2 = \left[ f(x) - f(N-x) \right]^2$ (i.e.\ $f(x) =0$ or $f(N-x)=0$). The latter is only true when $r=1$. By the assumption that $r<1,$ $\Trace{(\rhofx)^2} - (\Trace{\rhofx})^2 \neq 0$ and Proposition~1 gives that $\scorex$ derives from Eq~\eqref{eq:sldtwobytwosecond}. This together with Eq.~\eqref{eq:quantumfishertwo} yields
\begin{widetext}
\begin{equation}
	H(\lambda) = 4 m^2 (1-2\lambda)^{2m-2}
	             \sum_{x=0}^{(N-1)/2} 
							 \frac{\left[ f(x) - f(N-x)\right]^2 \left[ f(x) + f(N-x)\right]}
							      {\left[ f(x) + f(N-x)\right]^2
										 - 
										 (1-2\lambda)^{2m}\;
										 \left[ f(x) - f(N-x)\right]^2
							      }.
	\label{eq:genphaseflipqfione}
\end{equation}
\end{widetext}
The term within the summation of Eq.~\eqref{eq:genphaseflipqfione} yields the same quantity when $x$ is replaced by $N-x.$ Thus the range of the sum can be extended to $0$ to $N$ provided that the result is halved. Also the term within the summation only depends on the number of zeroes in the bit string for $x$ and different $x$'s can give the same contribution. Let $j$ denote a possible number of zeroes. For $x$ ranging from $0$ to $N$ the number of terms whose bit strings have $j$ zeroes is $\binom{n}{j}.$ For each of these terms $f(x) \pm f(N-x) = [(1+r)^j(1-r)^{n-j} \pm (1+r)^{n-j}(1-r)^j ] /2^n.$ Thus
\begin{eqnarray}
	H(\lambda) & = & \frac{m^2 (1-2\lambda)^{2m-2}}{2^{n-1}}
	                 \nonumber \\
						 & & \times
	             \sum_{j=0}^{n} 
							 \binom{n}{j}
							 \frac{c_j^2 d_j}{d_j^2 - (1-2\lambda)^{2m} c_j^2}.
	\label{eq:genphaseflipqfitwo}
\end{eqnarray}
where
\begin{subequations}
 \label{eq:partsofH}
 \begin{eqnarray}
   c_j & = & (1+r)^j(1-r)^{n-j} \nonumber \\
	       &    & - (1+r)^{n-j}(1-r)^j  \\
	       d_j & = &  (1+r)^j(1-r)^{n-j} \nonumber \\
	       &    &+ (1+r)^{n-j}(1-r)^j. 
 \end{eqnarray}
\end{subequations}
Equations~\eqref{eq:genphaseflipqfitwo} and~\eqref{eq:partsofH} form the central result from which the principal conclusions of this  article follow.

The estimation accuracies of this correlated state protocol and the optimal independent channel use protocol can be compared using the quantum Fisher informations of Eqs.~\eqref{eq:genphaseflipqfitwo} and~\eqref{eq:Hmanyindepoptimal}. The constraints are that \emph{the polarizations, channel parameter values and number of channel invocations are identical for each protocol respectively}. Given this, the fractional gain provided by the correlated state protocol is 
\begin{equation}
	G(\lambda):= \frac{H(\lambda)}{\Hoptind}
\end{equation}
and when $G(\lambda) >1,$ the correlated state protocol performs more accurately than the independent channel use protocol.
Equations~\eqref{eq:Hmanyindepoptimal} and~\eqref{eq:genphaseflipqfitwo} give
\begin{eqnarray}
	G(\lambda) & = &\frac{m(1-2\lambda)^{2m-2}}{2^{n+1}r^2}\;
	        \nonumber \\
		 & & \times
	     \sum_{j=0}^{n} 
				 \binom{n}{j}
				 \frac{c_j^2}{d_j}\;
				 \frac{1 - (1-2\lambda)^2 r^2}{1 - (1-2\lambda)^{2m} c_j^2/d_j^2}.
	\label{eq:gengain}
\end{eqnarray}

For a fixed polarization $0 \leqslant r <1$ the gain is a monotonically increasing function of $(1-2\lambda)^2.$ This follows from the fact that the gain function is a sum of terms of the form $(1/x - r^2)/(1/x^m - c_j^2/d_j^2)$ with $x=(1-2\lambda)^2.$ Then, as shown in appendix~\ref{app:Hbound}, $c_j^2/d_j^2 \geqslant r^2$ (the only exception occurs when $j=n/2$, in which case $c_j=0$ and this term is irrelevant in the sum). Thus as $x$ increases from $0$ to $1$, the fraction $(1/x - r^2)/(1/x^m - c_j^2/d_j^2)$ increases. It follows that \emph{for a fixed polarization the minimum gain is}
\begin{equation}
  G_\textrm{min} = 
	\begin{cases}
	\displaystyle
	\frac{1}{2^{n+1}r^2}\;
	     \sum_{j=0}^{n} 
				 \binom{n}{j}
				 \frac{c_j^2}{d_j} & \textrm{if $m=1$} \\
	0 & \textrm{if $m \neq 1$}
	\end{cases}
	\label{eq:mingengain}
\end{equation}
and in each case these are attained when $\lambda = 1/2.$ Similarly \emph{for a fixed polarization the maximum gain is}

\begin{equation}
	G_\textrm{max} = \frac{m}{2^{n+1}r^2}\;
	     \sum_{j=0}^{n} 
				 \binom{n}{j}
				 \frac{c_j^2}{d_j}\;
				 \frac{1-r^2}{1- c_j^2/d_j^2}.
	\label{eq:maxgengain}
\end{equation}
and this is attained when $\lambda =0\: \textrm{or} \: 1.$ 

The other extremes of the region of parameter space are $r \rightarrow 0$ and $r \rightarrow 1.$ In the first case, equivalent to $r \ll 1,$ 
$c_j \approx 2r(2j-n)$ and $d_j \approx 2.$ Eq.~\eqref{eq:gengain} then implies that
\begin{equation} 
  G \rightarrow mn(1-2\lambda)^{2m-2} \quad \textrm{as $r \rightarrow 0$}.
	\label{eq:gengainrzero}
\end{equation}
As $r \rightarrow 1$, $c_j \rightarrow 0$ unless $j=0 \: \textrm{or} \: n.$ Specifically $c_0 = -c_n \rightarrow -2^n$ and $d_0=d_n \rightarrow 2^n.$ Thus Eq.~\eqref{eq:gengain} implies that 
\begin{equation} 
  G \rightarrow m (1-2\lambda)^{2m-2}\frac{1- (1-2\lambda)^2}{1- (1-2\lambda)^{2m}} \quad \textrm{as $r \rightarrow 1$}.
	\label{eq:gengainrone}
\end{equation}
Note that this excludes the cases where both $\lambda =0 \: \textrm{or}\: 1$ and $r=1$ here as Proposition~1 will give an SLD and Fisher information different to that of Eq.~\eqref{eq:genphaseflipqfione}. These cases are beyond consideration here.  

The preceding results indicate important differences between situations in which the phase-flip operation is invoked only once and those where it is invoked more than once.



\section{Estimation with a single channel use}
\label{sec:correstone}

If the phase-flip is invoked only once then the correlated state protocol yields the following gain
\begin{equation}
	G(\lambda) = \frac{1}{2^{n+1}r^2}\;
	     \sum_{j=0}^{n} 
				 \binom{n}{j}
				 \frac{c_j^2}{d_j}\;
				 \frac{1 - (1-2\lambda)^2 r^2}{1 - (1-2\lambda)^2 c_j^2/d_j^2}.
	\label{eq:gain}
\end{equation}
As shown in appendix~\ref{app:Hbound}, with just one phase-flip operation and for any $n \geqslant 2,$ $G(\lambda) > 1$ if $ 0 < r < 1$. This shows that, \emph{for single phase-flip invocation, the correlated state protocol of section~\ref{sec:correst} gives a more accurate estimate of the phase-flip parameter than the independent channel use protocol whenever the polarizations used by each are identical.} Additionally Eqs.~\eqref{eq:gengainrone} and~\eqref{eq:gengainrzero} imply that $G \rightarrow 1$ as $r \rightarrow 1$ and that $G \rightarrow n$ as $r \rightarrow 0.$ The striking observation is that \emph{for sufficiently small polarizations and a single phase-flip invocation, the correlated state protocol increases the estimation accuracy by a factor of approximately $n$ regardless of the value of the parameter.}

For polarization values that do not approach these limits, a relevant measure of success would be the minimum gain, which is, according to Eq.~\eqref{eq:mingengain},
\begin{equation}
	G_\textrm{min} = 
	\frac{1}{2^{n+1}r^2}\;
	     \sum_{j=0}^{n} 
				 \binom{n}{j}
				 \frac{c_j^2}{d_j} 
\end{equation}
and is larger than $1$. 

\begin{figure}[ht]
  \includegraphics[scale=.80]{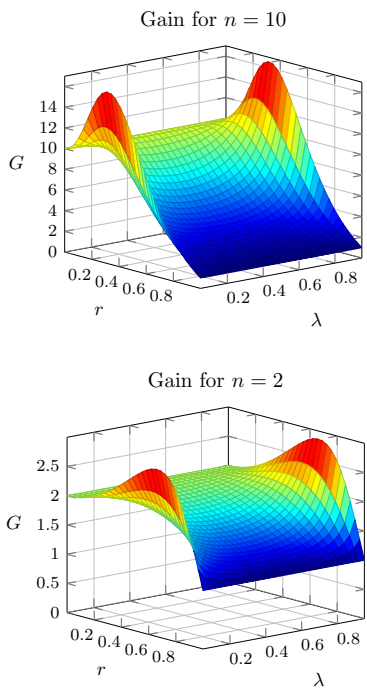} 
  \caption{Gain using one phase-flip invocation. Both plots are over the range $0.05 \leqslant \lambda \leqslant 0.95$ and $ 
	0 \leqslant r \leqslant 1.$
           \label{fig:gainplotssingle}}
 \end{figure}

Typical plots of the gain, which illustrate the features described above, are provided in Fig.~\ref{fig:gainplotssingle}.
These indicate that for  a fixed value of $\lambda$, the gain does not necessarily increase or decrease with increasing value of $r,$ although it appears that $G_\textrm{min}$ decreases as $r$ increases. Also, depending on the polarization and parameter value, it is possible to attain gains in excess of $n$. For example, with $n=2$ the extremes are $G_\textrm{min} = 2/(1+r^2)$ and $G_\textrm{max} = 2(1+r^2)/(1-r^2).$ Clearly a gain of arbitrarily large magnitude  can be attained provided that, for example, $\lambda$ and $r$ are sufficiently close to $1$.



\section{Estimation with multiple channel uses}
\label{sec:correstmany}

In the more general case of the protocol of Figs.~\ref{fig:preparedscheme} and~\ref{fig:uprep}, there are $m$ phase-flip invocations on $n$ qubits.  Here Eq.~\eqref{eq:gengain} indicates that $G(1/2)=0$ and thus there will be ranges of phase-flip parameters for which the correlated state protocol provides a gain of less than $1$ and thus gives a lower accuracy than the independent channel use protocol. 

\begin{figure}[ht]
  \includegraphics[scale=.80]{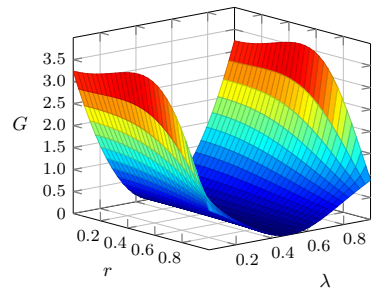} 
  \caption{Gain using two phase-flip invocations on two qubits, plotted for $0.05 \leqslant \lambda \leqslant 0.95$ and $ 
	0 \leqslant r \leqslant 1.$
           \label{fig:gainplotstwo}}
 \end{figure}

A relatively simple example has $n=2$ and $m=2$ and the gain is illustrated in Fig~\ref{fig:gainplotstwo}.
This indicates regions in the parameter space for which $G(\lambda) >1$ and other regions for which $G(\lambda) <1.$ Comparing Figs.~\ref{fig:gainplotssingle} and~\ref{fig:gainplotstwo} for $n=2$ indicate regions for values of $\lambda$ near to $0$ or $1$ in which the gain using two phase-flip invocations is larger than that using one phase-flip invocation. 

\begin{figure}[ht]
  \includegraphics[scale=.80]{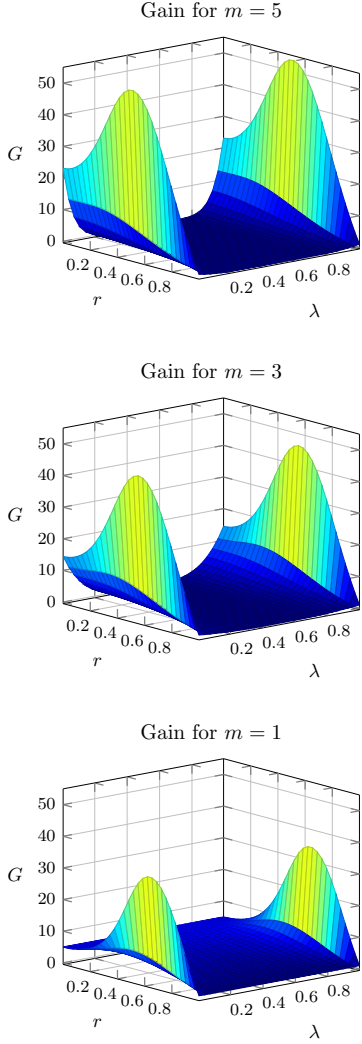} 
  \caption{Gain using various numbers of  phase-flip invocations on five qubits, plotted for $0.0005 \leqslant \lambda \leqslant 0.9995$ and $0 \leqslant r \leqslant 1.$
           \label{fig:gainplotsmany}}
 \end{figure}

Whenever $m \geqslant 2,$ Eq.~\eqref{eq:mingengain} precludes any advantages for the correlated state protocol over the \emph{entire range} of $\lambda.$ Similarly Eq.~\eqref{eq:gengainrzero} results in the same loss of advantage over the entire parameter range when $r \rightarrow 0.$ However, Eq.~\eqref{eq:maxgengain} shows that as $\lambda \rightarrow 0$ or $\lambda \rightarrow 1$ the gain approaches $m$ times that of the correlated state protocol which uses only one phase-flip invocation. Thus for such parameter values the correlated state protocol will be superior to the independent channel use protocol. These facts are illustrated in Fig.~\ref{fig:gainplotsmany}. 

It is possible to attain a substantial gain. For example, if $r \ll 1$, then Eq.~\eqref{eq:gengainrzero} shows that $G(\lambda) \geqslant n$ provided that 
\begin{equation}
	\lambda \leqslant \frac{1}{2}\; \left[ 1 - \frac{1}{m^{1/(2m-2)}} \right].
\end{equation}
The right hand side of this decreases as $m$ increases. Note that a symmetrical situation occurs by replacing $\lambda$ by $1-\lambda.$

\begin{figure}[ht]
  \includegraphics[scale=.80]{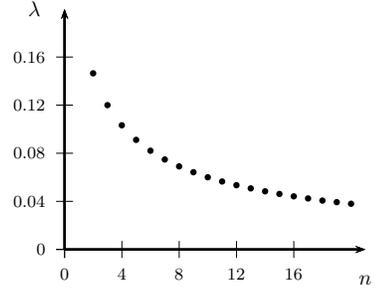} 
  \caption{Maximum value of $\lambda$ vs $n$ such that $G(\lambda)\geqslant n$ for cases where $r \ll 1.$
           \label{fig:lambdabound}}
 \end{figure}

Of special interest are the cases where the phase-flip is applied to all qubits (i.e.\ $m=n$).  Here, the resulting upper bounds on $\lambda$ such that $G(\lambda) \geqslant n$ are illustrated in Fig.~\ref{fig:lambdabound}.
This situation appears in NMR implementations of quantum information processing, where all of the qubits are subjected to the same or similar phase-flip operations. In room-temperature solution-state NMR implementations~\cite{nielsen00}, the polarization of a single qubit is typically $r \approx 10^{-4} $, which satisfies the assumptions resulting in Fig.~\ref{fig:lambdabound}. Considering, $\lambda =0.10$, a typical value appearing in Fig.~\ref{fig:lambdabound}, Eq.~\eqref{eq:lambdaT2} shows that $\lambda  \leqslant 0.10$ when $t  \leqslant 0.22 T_2.$ In typical NMR quantum information processing experiments~\cite{collins00}, $T_2 \approx 1\,\textrm{s}$ and thus by limiting the period of evolution of the system to less than about $0.2\,\textrm{s},$ the gain attained in estimating $\lambda$ will exceed $n$. This is certainly feasible to attain.



\section{Entanglement and Discord Considerations}
\label{sec:entangdisc}

It is customary to inquire about specifically which quantum resources could be responsible for gains of the type provided by the correlated state protocol. In other instances where quantum systems offer gains for parameter estimation, attention has focused on the presence of entanglement or other types of quantum correlations~\cite{giovannetti06,modi11}. In the context of mixed state phase estimation it was shown that entanglement was not necessary for improved accuracy but it appears that quantum discord, a more general form of quantum correlation, is present whenever there is a gain in accuracy~\cite{modi11}.

The preparation scheme for Fig.~\ref{fig:uprep} is capable of producing states for which the measurement outcomes are correlated and it is of interest to determine the extent to which various quantum correlations are present in our scheme.  The simplest context in which to address this for the correlated state protocol is that involving two qubits. For $n=2$ the gain is
%
\begin{eqnarray}
	G(\lambda) & = & 2m(1-2\lambda)^{2m-2}\;(1+r^2)\; \nonumber \\
	                  & &  \times
	                     \frac{1-(1-2\lambda)^2r^2}{(1+r^2)^2 -4r^2(1-2\lambda)^{2m}}.
											\label{eq:gainfortwo}
\end{eqnarray}
%
We aim to determine whether the presence of entanglement is necessary for $G(\lambda) >1$ and whether other types of quantum correlation may be responsible for the gain attained by the correlated state protocol. We offer a preliminary investigation of these issues, focusing mostly on the presence of these correlations. A more complete treatment, which may consider issues such as the dependence of discord on the purity of the state and the extent of classical correlations in these protocols, is beyond the scope of this article.

The existence of entanglement or other forms of quantum correlation is inferred from the density operator for the system. After $m$ channel invocations, Eq.~\eqref{eq:rhofinalafterm} yields that, in the basis $\left\{ \ket{00}, \ket{01}, \ket{10},\ket{11}\right\}$,
\begin{equation}
	\rhof = \frac{1}{2}
	        \begin{pmatrix}
	         \dfrac{1+r^2}{2} & 0 & 0 & i r\mu \\
					 0 & \dfrac{1-r^2}{2} & 0 & 0\\
					 0 & 0 &\dfrac{1-r^2}{2} & 0 \\
					 -i r\mu & 0 & 0 & \dfrac{1+r^2}{2}
	        \end{pmatrix}.
	\label{eq:finalrhofortwo}
\end{equation}
where $\mu := (1-2\lambda)^m.$ Note that the system density operator prior to channel invocation is the same as that for which $\lambda=0,$ or, equivalently, $\mu=1.$

\subsection{Entanglement and separability}
\label{sec:entang}

A necessary and sufficient condition for a two qubit state to be separable (non-entangled) is that the partial transpose of the density operator for the state is a positive operator~\cite{peres96,horodecki96}.  The partial transpose of $\rhof$ is
\begin{equation}
	\rhof^\mathrm{PT} = \frac{1}{2}
	        \begin{pmatrix}
	         \dfrac{1+r^2}{2} & 0 & 0 &  0 \\
					 0 & \dfrac{1-r^2}{2} & i r\mu & 0\\
					 0 & -i r\mu &\dfrac{1-r^2}{2} & 0 \\
					 0 & 0 & 0 & \dfrac{1+r^2}{2}
	        \end{pmatrix}.
\end{equation}
This will be a positive operator if and only if its eigenvalues are all positive. The block diagonal form of this matrix renders the computation of the eigenvalues straightforward and it can be shown that these are positive if and only if 
$ 0< r < \sqrt{(1-2\lambda)^{2m} + 1} - \left| (1-2\lambda)^{m}\right|.$
Thus there will be polarizations such that for all $\lambda$ and $m$, $\rhof$ is separable. Yet the results of Sections~\ref{sec:correstone} and~\ref{sec:correstmany} indicate the correlated state protocol yields greater accuracy than the independent channel use protocol for arbitrarily small polarizations and when $m=1$. Thus \emph{the gains of the correlated state protocol cannot be ascribed to entanglement.} 

\subsection{Discord}
\label{sec:disc}

Quantum discord is an alternative and more general characterization of the correlations between two quantum systems~\cite{ollivier01,modi11a}. Discord requires the density operator for the bipartite system, $\rhoop$ and the reduced density operators for the two the subsystems $\rhoop^\textrm{A}$ and $\rhoop^\textrm{B}$, each of which is attained by taking the partial trace over the complementary system. The  quantum mutual information is ${\cal I}(\rhoop):= S(\rhoop^\textrm{A}) + S(\rhoop^\textrm{B}) - S(\rhoop)$ where the von Neumann entropy is $S(\rhoop) : = - \Trace(\rhoop \log_2{\rhoop}).$ 

A possible alternative approach to defining the quantum mutual information supposes that a measurement, described by POVM operators $\{ \hat{E}_k \} $  where $k$ indexes the possible measurement outcomes, has been performed on system B and then inquires about the remaining information in subsystem A. Outcome $k$ occurs with probability $p_k := \Trace{(\hat{E}_k \rhoop)}$ and the state of system A following this outcome is $\rhoop_{A \vert k}:= \Trace_B{(\rhoop \hat{E}_k)}/p_k.$  If the systems were in a product state prior to this measurement, then  $\rhoop^\textrm{A} = \rhoop_{A \vert k}$ for all $k$.  Thus a measure of the classical correlation between the subsystems, conditional on this particular measurement, can be quantified by $S(\rhoop^\textrm{A}) - \sum_k p_k S(\rhoop_{A \vert k})$, where the sum is an average over all possible measurement outcomes. Let $S(\rhoop \vert \{ \hat{E}_k \} ) := \sum_k p_k S(\rhoop_{A \vert k})$ . The overall classical correlation between the subsystems is them an maximum over all possible measurements and is  ${\cal C}(\rhoop) := S(\rhoop^\textrm{A}) - \min {S}(\rhoop \vert \{ \hat{E}_k \} )$ where the minimization is done over all POVMs on the subsystem B. The quantum discord is defined as the difference between the quantum mutual information and the classical correlation, ${\cal Q}:= {\cal I}(\rhoop)  - {\cal C}(\rhoop) .$ In general, ${\cal Q}(\rhoop) \geqslant 0 $ and it is invariant under local unitary transformations on the subsystems~\cite{ollivier01,modi11a}.

The requirement for minimization over all possible measurements means that computing the quantum discord is generally very difficult. However, analytical expressions exist for certain classes of two-qubit states~\cite{luo08,chen11}. In particular an expression exists for for ``X-states,'' which have the form, 
\begin{equation}
	\rhoop = \frac{1}{4}\;
	              \left( 
								 \hat{I} 
								 +\sum_{j=1}^3 c_j \hat{\sigma}_j \otimes \hat{\sigma}_j
								\right)
	\label{eq:rhoforQ}
\end{equation}
where $c_j$ are all real and $ \hat{\sigma}_1 = \sigmax, \ldots. $ Let 
\begin{subequations}
 \label{eq:evalsforQ}
\begin{eqnarray}
 \lambda_0 & := & 1-c_1-c_2-c_3 \\
 \lambda_1 & := &  1-c_1+c_2+c_3  \\
 \lambda_2 & := & 1+c_1-c_2+c_3  \\
 \lambda_3 & := & 1+c_1+c_2-c_3 
\end{eqnarray}
\end{subequations}
and 
\begin{equation}
	c:= \max{ \left\{ \left| c_1\right|, \left| c_2\right|,\left| c_3\right| \right\}.
	             }
\end{equation}
Then~\cite{luo08} 
\begin{eqnarray}
	{\cal Q}(\rhoop) & = & \frac{1}{4}\sum_{j=0}^3 \lambda_j \log_2{\lambda_j}  -\frac{1-c}{2}\;  \log_2{\left( 1-c \right) }  \nonumber \\
	                         & &  - \frac{1+c}{2}\;  \log_2{\left( 1+c \right) }.
\end{eqnarray}

The final density operator of Eq.~\eqref{eq:finalrhofortwo} can be brought into the form (as it is the relevant values of $c_j$ would not all be real) of Eq.~\eqref{eq:rhoforQ} by the single qubit unitary
\begin{equation}
 \hat{U} = 
 \begin{pmatrix}
   0 & e^{i\pi/8} \\
	  e^{-i\pi/8}  & 0
 \end{pmatrix}	
\end{equation}
applied to each qubit. This will not alter the quantum discord. After this, direct calculation gives
\begin{eqnarray}
 \lambda_0 & := & 1-r^2 \\
 \lambda_1 & := &  1+ 2r(1-2\lambda)^m  + r^2 \\
 \lambda_2 & := & 1- 2r(1-2\lambda)^m  + r^2  \\
 \lambda_3 & := & 1-r^2 
\end{eqnarray}
and 
\begin{equation}
	c:= \max{ \left\{r^2, r(1-2\lambda)^m \right\}.
	             }
\end{equation}
The discord for $\rhoprep$ can be determined by setting $\lambda=0$ in these expressions. Then $c=r$ and 
\begin{equation}
	{\cal Q}(\rhoop) = \frac{1+r}{2}\;
	                             \log_2{\left( 1+r \right) } 
															+ \frac{1-r}{2}\;
	                             \log_2{\left( 1-r \right) }.
\end{equation}
If $r>0,$ then $\frac{\partial {\cal Q}}{\partial r} >0$ and, since ${\cal Q} =0$ when $r=0,$ this implies that the system state prior to the channel invocation has non-zero quantum discord. In this sense, there are always quantum correlations present prior to channel invocation. 

After channel invocation, the discord depends on whether $r>\mu:= (1-2\lambda)^m$ or not. In all cases, it is clear that replacing $\mu$ with $-\mu$ does not alter ${\cal Q}$ and thus it suffices to consider $\mu >0$ (negative values are possible if $m=1$). In appendix~\ref{app:discordprop} it is shown that for $\mu >0$ 
\begin{equation}
	\frac{\partial{\cal Q}}{\partial \mu} >0 \;\; \textrm{and} \;\;
	\frac{\partial {\cal Q}}{\partial r} >0.
\end{equation}

With regard to $\lambda$, one extreme is $\lambda =1/2$ (i.e.\  $\mu =0$). Here $c= r^2$ and 
\begin{equation}
	{\cal Q}(\rhoop) =0.
\end{equation}
Thus regardless of the number of channel invocations there is no quantum discord when $\lambda =1/2$ (the final state of the system is clearly separable). However, with a single channel invocation, the gain in quantum Fisher information is always at least $1$ and approaches $2$ as $r \rightarrow 0.$ Thus \emph{the gains provided by the correlated state protocol cannot always be attributed to non-zero quantum discord.}

Beyond this extreme, i.e.\  $\mu >0,$ the discord is non-zero. For a fixed value of $\lambda \neq 0,$ the discord is a monotonically increasing function of $r$. However, Eq.~\eqref{eq:gainfortwo} can be used to show that the gain is not monotonically increasing. This requires finding $r$ such that $\frac{\partial G}{\partial r} =0$. Explicit calculation reveals that this occurs when $r=0$ or 
\begin{equation}
	r^2 = \frac{1+\nu \pm 2\sqrt{(1+\nu)^2\nu^m - 4 \nu^{m+1}}}{4 \nu^{2m+1} -1 -\nu}.
											\label{eq:gainfortwomax}
\end{equation}
where $\nu = (1-2\lambda)^2.$ Table~\ref{tab:gainmax}  lists examples where $r$ is such that $\frac{\partial G}{\partial r} =0$.

\begin{table}[h]
 \begin{tabular}{c|c|c}
   $m$ & $\lambda$ & $r$ \\
	 \hline
	 $1$ & $0.95$& $0.66$\\
	 \hline
	 $1$ & $0.99$& $0.83$ \\
	 \hline
	 $2$ & $0.95$& $0.48$\\
	 \hline
	 $2$ & $0.99$& $0.76$\\
 \end{tabular}
 \caption{Values of $r$ such that $\frac{\partial G}{\partial r} =0$ as calculated using Eq.~\eqref{eq:gainfortwomax}.
          \label{tab:gainmax}
					}
\end{table}

This indicates that the gain is not a monotonically increasing function of $r$. Thus there will be regions where the discord of the pre-measurement state increases while the gain attained decreases and \emph{an increase in the discord corresponds to a decrease in the gain}. This is also evident from the lower graph of Fig.~\ref{fig:gainplotssingle}.



\section{Conclusion}
\label{sec:summary}

We have considered two general phase-flip parameter estimation protocols and compared their accuracies as quantified by the quantum Fisher information. The first protocol, the independent state protocol, proceeds in such a way that the states of the system qubits are always non-entangled. This is akin to the classical strategy of repeating the estimation procedure independently multiple times and offers the same quantitative advantages. The second protocol, uses states which are correlated via an entangling unitary prior to channel invocation. If the number of channel invocations is identical and the polarizations are identical, we have shown that the correlated state protocol can improve on the accuracy of the parameter estimation when compared to the independent state protocol. The extent of the improvement depends on the actual parameter value but it can be substantially larger than the typical gain attained in the classical strategy. 

Our analysis demonstrates that the correlated state protocol attains advantages when the state of the system prior to measurement is separable and, in some cases,  where there is zero quantum discord. Also, we have shown that, in some situations, an increase in quantum discord is associated with a decrease in the gain in estimation accuracy. The gains afforded by the correlated state protocol cannot be attributed entirely to entanglement or non-zero quantum discord and it is unclear which quantum resources are possibly responsible for the gains. 

The correlated state protocol may be of use in estimating the dephasing time, $T_2$ of a spin in NMR as this is closely related to the phase-flip parameter, $\lambda.$ However, we have assumed that the phase-flip parameter is identical for all qubits; this is not necessarily true in implementations.  For example, NMR systems used in quantum information processing provide different dephasing times amongst the system qubits (for examples from NMR quantum information processing see~\cite{marx00,das02,peng02,ramanathan04,xiao06}). It remains to be seen what gains methods like those of this article will yield in such cases. 

We stress that the correlated state protocol that we analyzed uses one particular preparatory unitary, $\uprep,$ and we have not determined whether this is optimal. 

Finally, the methods used here are applicable to other Pauli-type channels, i.e.\ where $\sigmaz$ is replaced by $\sigma_n$ with $n$ corresponding to any direction, in Eq.~\eqref{eq:phaseflip}. These channels are all related by single qubit rotations and the latter can be absorbed into the various parameter independent unitaries and measurements used in the estimation protocol.


\acknowledgments

The author would like to thank Michael Frey for many useful discussions.

\appendix


\section{State after preparatory unitary}
\label{app:rhoprep}

The Hadamard gate of Fig.~\ref{fig:uprep}  acts on a single qubit as
\begin{equation}
	\ket{z_j} \rightarrow \frac{1}{\sqrt{2}}\; \sum_{y_j=0}^1 \left( -1 \right)^{z_j y_j}\; \ket{y_j}
\end{equation}
and the controlled-$Z$ between bits $j$ and $k$ via
\begin{equation}
	\ket{z_n \ldots z_1} \rightarrow  \left( -1 \right)^{z_j z_k}\; \ket{z_n \ldots z_1}.
\end{equation}
Thus 
\begin{equation}
 \uprep = \frac{1}{2^{n/2}}\;
          \sum_{y,z} \left( -1 \right)^{z \cdot y + s}
					\ket{y}\bra{z}
 \label{eq:upreprep}
\end{equation}
where the binary representation of $z$ is $z_n\ldots z_1$ and the sum is over all possible values of these binary representations of $z$ and $y$. Also $z \cdot y := \sum_{j=1}^n z_j y_j$ and $s:= z_1z_2 + z_1z_3 + \ldots + z_{n-1}z_n$, in which each distinct pair of non-equal subscripts appears exactly once. 

In order to determine $\rhoprep : =  \uprep \rhoi \uprep^\dagger,$ it will be convenient to re-express $\rhoi = \rhoo \otimes \rhoo \otimes \cdots \otimes \rhoo$ in terms of the single qubit basis states, $\ket{\pm} = (\ket{0} \pm \ket{1})/\sqrt{2}.$ Here 
\begin{equation}
	\rhoo = q \ket{+}\bra{+} + (1-q) \ket{-}\bra{-}
	\label{eq:rhhobasis}
\end{equation}
where $q=(1+r)/2.$ A convenient representation of $\rhoi$ is in terms of a sum of mutually orthogonal operators, each of which has a two dimensional support. Thus
\begin{equation}
	\rhoi = \sum_{x=0}^{(N-1)/2} \rhoix
	\label{eq:rhoinitdecomp}
\end{equation}
where straightforward algebra using Eq~\eqref{eq:rhhobasis} gives
\begin{equation}
	\rhoix = f(x) \ket{\overline{x}}\bra{\overline{x}} + f(N-x) \ket{\overline{N-x}}\bra{\overline{N-x}}
	\label{eq:rhoinitcomponent}
\end{equation}
with $f$ as defined in Eq.~\eqref{eq:fdefn}.
Here $\ket{\overline{x}}$ is such that if $x_n \ldots x_1$ is the binary representation of $x$, then $\ket{\overline{x}}$ contains the binary string for $x$ with ``$0$'' replaced by ``$+$'' and ``$1$'' replaced by ``$-$''. For example with $n=3$, $\ket{\overline{0}} = \ket{+++} = \ket{+}\ket{+}\ket{+}, \ket{\overline{1}}= \ket{++-}, \ket{\overline{2}} =\ket{+-+},$ etc,\ldots. In this example $\rho_\textrm{i (0)}$ has support spanned by $\{ \ket{+++}, \ket{---} \}$, $\rho_\textrm{i (1)}$ has support spanned by $\{ \ket{++-}, \ket{--+} \}$ and so on.

The preparatory unitary results in 
\begin{equation}
	\rhoprep = \sum_{k=0}^{(N-1)/2} \rhox
	\label{eq:rhoinitdecomptwo}
\end{equation}
where 
\begin{eqnarray}
	\rhox & := & \uprep \rhoix \uprep^\dagger
	             \nonumber \\
        & = & f(x) \uprep \ket{\overline{x}}\bra{\overline{x}} \uprep^\dagger
				      \nonumber \\
				& &  + f(N-x) \uprep \ket{\overline{N-x}}\bra{\overline{N-x}} \uprep^\dagger.
	\label{eq:rhoicompexpanded}
\end{eqnarray}
 Each term in the decomposition of Eq.~\eqref{eq:rhoinitdecomptwo} is an operator with two dimensional support and these supports are all mutually orthogonal. Specifically the support of $\rhox$ is $\{ \uprep \ket{\overline{x}}, \uprep \ket{\overline{N-x}} \}$ and the unitary nature of the preparatory gate results in the orthogonality of distinct supports. 

It remains to determine an expression for $\uprep \ket{\overline{x}}$ in terms of the computational basis (i.e.\ with bits strings consisting of ``$0$'' and ``$1$'' terms rather than ``$+$'' or ``$-$'' terms). For a single qubit
\begin{equation}
	\ket{\overline{x_j}} = \frac{1}{\sqrt{2}}\; 
	                       \sum_{z_j=0}^1
												 \left( i \right)^{z_j} \left( -1 \right)^{x_j z_j}
												 \ket{z_j}
\end{equation}
and this gives
\begin{equation}
	\ket{\overline{x}} = \frac{1}{2^{n/2}}\; 
	                       \sum_{z=0}^{N}
												 \left( i \right)^{z_1 + \ldots + z_n} \left( -1 \right)^{x \cdot z}
												 \ket{z}.
	\label{eq:ybasisrep}
\end{equation}
Combining Eqs.~\eqref{eq:upreprep} and~\eqref{eq:ybasisrep} gives
\begin{equation}
	\uprep \ket{\overline{x}} = \frac{1}{2^n}\;
	                            \sum_{y,z =0}^{N}
															\left( i \right)^{z_1 + \ldots + z_n}
															\left( -1 \right)^{y\cdot z + x \cdot z + s}
															\ket{y}
	\label{eq:upreponstate}
\end{equation}
where $s:= z_1z_2 + z_1z_3 + \ldots + z_{n-1}z_n$, in which each distinct pair of non-equal subscripts appears exactly once. We shall show that 
\begin{equation}
	\uprep \ket{\overline{x}} = \frac{1+i}{2}\; \ket{x}
	                            + \frac{1-i}{2}\; \ket{N-x}
  \label{eq:upreponstatesimple}
\end{equation}
(recall that $\ket{x}$ contains the bit string for $x$ in terms of ``$0$'' and ``$1$'' while $\ket{\overline{x}}$ contains the bit string for $x$ in terms of ``$+$'' and ``$-$''). Substitution into Eq.~\eqref{eq:rhoicompexpanded} then yields the result of Eq.~\eqref{eq:rhoprepped}.

In order to show Eq~\eqref{eq:upreponstatesimple}, consider $\bra{x} \uprep \ket{\overline{x}}$. Then Eq.~\eqref{eq:upreponstate} gives
\begin{eqnarray}
	\bra{x} \uprep \ket{\overline{x}} & = & \frac{1}{2^n}\; 
	                            \sum_{z =0}^{N}
															\left( i \right)^{z_1 + \ldots + z_n}
															\left( -1 \right)^{x\cdot z + x \cdot z + s}
															\nonumber \\
															& = & \frac{1}{2^n}\; 
	                            \sum_{z =0}^{N}
															\left( i \right)^{z_1 + \ldots + z_n}
															\left( -1 \right)^s
	\label{eq:keysum}
\end{eqnarray}
since $\left( -1 \right)^{x\cdot z + x \cdot z} = 1.$ The term in the sum, $\left( i \right)^{z_1 + \ldots + z_n} \left( -1 \right)^s$  depends only on the number, $m$, of $z_1, \ldots z_n$ which are equal to $1.$  Clearly $\gamma:=\left( i \right)^{z_1 + \ldots + z_n} = i^m.$  The only bits which contribute to $s$ are those for which \emph{both}  $z_j = 1$ and $z_k =1$. Given that $m$ of $z_1, \ldots z_n$ are equal to $1$, the number of such such pairs is $\binom{m}{2} = m(m-1)/2$. Depending on whether this is even or odd, the term $\beta:= \left( -1 \right)^s$ gives $\pm 1.$  The various possibilities for $m$ and their contributions are summarized in Table~\ref{tab:summands}.

\begin{table}[h]
 \begin{tabular}{c|c|c|c}
   $m$ & $\beta$ & $\gamma$ & $\beta\gamma$ \\
	 \hline
	 $4k$ & $1$& $1$& $1$\\
	 \hline
	 $4k+1$ & $1$& $i$& $i$\\
	 \hline
	 $4k+2$ & $-1$& $-1$& $1$\\
	 \hline
	 $4k+3$ & $-1$& $-i$& $i$\\
 \end{tabular}
 \caption{Contributions to the sum of Eq.~\eqref{eq:keysum}. Here $k=0,1,\ldots$ is an integer.
          \label{tab:summands}
					}
\end{table}

 This shows that whenever $m$ is even the contribution to the sum is $\beta \gamma = 1$ and whenever it is odd, $\beta \gamma =i.$ As $z$ ranges through all possible values, the number of times that $m$ is even is $2^n/2$ (exactly half of all bit strings have even numbers of bits equal to $1$). Thus there are $2^n/2$ contributions of $1$ and $2^n/2$ contributions of $i$. Thus
\begin{equation}
	\bra{x} \uprep \ket{\overline{x}} = \frac{1+i}{2}.
\end{equation}
Now consider $\bra{N-x} \uprep \ket{\overline{x}}$. Note that each bit in $\bra{N-x}$ is the bit flip of the corresponding bit in $\bra{x}$. Then in Eq.~\eqref{eq:upreponstate} 
\begin{equation}
	\bra{N-x} \left( -1 \right)^{y\cdot z}
															\ket{y}
						= \left( -1 \right)^{x\cdot z + z_1 + \ldots + z_n}.
\end{equation}
Thus
\begin{equation}
	\bra{N-x} \uprep \ket{\overline{x}} = \frac{1}{2^n}\; 
	                            \sum_{z =0}^{N}
															\left( - i \right)^{z_1 + \ldots + z_n}
															\left( -1 \right)^s
\end{equation}
which shows that $\bra{N-x} \uprep \ket{\overline{x}} = \left( \bra{x} \uprep \ket{\overline{x}} \right)^* = (1-i)/2.$ This determines the two coefficients that appear in Eq.~\eqref{eq:upreponstatesimple}. Since $\uprep \ket{\overline{x}}$ has unit modulus, there cannot be any other components. This demonstrates Eq.~\eqref{eq:upreponstatesimple} and proves Eq.~\eqref{eq:rhoprepped}.



\section{Bounds on $G(\lambda)$}
\label{app:Hbound}

The general form of the gain is given by Eq.~\eqref{eq:gengain}, which contains the term $c^2_j/d^2_j.$ This enters into various bounds on the gain and following is a key result.

\textit{Proposition 3:} Suppose that $r<1.$ If $j \neq n/2$ then
\begin{equation}
	\frac{c_j^2}{d_j^2} \geqslant r^2
\end{equation}
with equality if and only if $r=0.$ If $j = n/2$ then 
\begin{equation}
	\frac{c_j^2}{d_j^2} =0.
\end{equation}

\textit{Proof:} Clearly if $r=0$, then $c_j =0$ and thus $c_j/d_j = r.$ Thus assume that $r>0.$ Let 
								\begin{equation}
									f_j:= \frac{c_j}{d_j} \frac{1}{r}.
								\end{equation}
								By the definition of $c_j$ and $d_j,$ it follows that $f_{n-j} = - f_j.$ Thus it suffices to prove the result for $j \geqslant n/2.$
								
Two special cases present themselves. First, if $n$ is even and $j = n/2,$ Eq~\eqref{eq:partsofH} gives that $c_j=0.$ Second, if $n$ is odd and $j = (n+1)/2,$ then let $n=2l+1$. Thus $j=l+1$ and direct substitution into Eq~\eqref{eq:partsofH} gives $c_j/d_j = r.$ 
								
For all other cases $j>(n+1)/2$. Here
\begin{equation}
	f_j - 1= \frac{d_j(1-r)- 2(1+r)^{n-j}(1-r)^j}{d_jr}.
	\label{eq:frelation}
\end{equation}
The numerator of this expression gives
\begin{widetext}
\begin{eqnarray}
 d_j(1-r)- 2(1+r)^{n-j}(1-r)^j & = & (1+r)^j(1-r)^{n-j+1} + (1+r)^{n-j}(1-r)^{j+1} 
                                     \nonumber \\
															 &   & - 2(1+r)^{n-j}(1-r)^j
                                     \nonumber \\
															 & = & (1+r)^j(1-r)^{n-j+1} - (1+r)^{n-j +1}(1-r)^{j+1}.
\end{eqnarray}
\end{widetext}
Now if $0 \leqslant r<1$, then $\left[ (1+r)/(1-r) \right]^{2j - 1 -n} \geqslant 1$ provided that $j > (n+1)/2$ (note that equality occurs if and only if $r=0.$) Thus $(1+r)^{2j - 1 -n} \geqslant (1-r)^{2j - 1 -n}$. This implies that $(1+r)^j(1-r)^{n-j+1} - (1+r)^{n-j +1}(1-r)^{j+1} \geqslant 0,$ with equality if and only if $r=0.$. Furthermore, $d_j >0$ and for $r>0$, Eq.~\eqref{eq:frelation} gives that $f_j -1  > 0.$ This proves the proposition. \hspace{\fill}$\Box$

It follows from proposition~3 that, if $r<1$ then the summand in Eq.~\eqref{eq:gengain} satisfies
\begin{equation}
         \frac{c_j^2}{d_j}\;
				 \frac{1 - (1-2\lambda)^2 r^2}{1 - (1-2\lambda)^{2m} c_j^2/d_j^2} \geqslant
         \frac{c_j^2}{d_j}\;
				  \frac{1 - (1-2\lambda)^2 r^2}{1 - (1-2\lambda)^{2m} r^2}
\end{equation}
with equality only possible if $r=0$ (note that for $j=n/2$, the result of proposition~3 is irrelevant since $c_j=0$). Thus
\begin{eqnarray}
	G(\lambda) & \geqslant &\frac{m(1-2\lambda)^{2m-2}}{2^{n+1}r^2}\;
	        \nonumber \\
		 & & \times
				 \frac{1 - (1-2\lambda)^2 r^2}{1 - (1-2\lambda)^{2m} r^2}\;
	     \sum_{j=0}^{n} 
				 \binom{n}{j}
				 \frac{c_j^2}{d_j}.
	\label{eq:gengainapprox}
\end{eqnarray}

In this context, the following is useful.

\textit{Proposition 4:} If $n \geqslant 2$ then
\begin{equation}
	\sum_{j=0}^{n} 
				 \binom{n}{j}
				 \frac{c_j^2}{d_j} \geqslant 2^{n+1} r^2
\end{equation}
with equality if $r=0. $

\textit{Proof:} If $r=0$ then $c_j=0$ and the equality holds. In general, straightforward algebra yields
\begin{equation}
	c_j^2 = d_j^2 - 4 (1-r^2)^n.
\end{equation}
Thus
\begin{eqnarray}
	\sum_{j=0}^{n} 
				 \binom{n}{j}
				 \frac{c_j^2}{d_j} & = &   \sum_{j=0}^{n} 
				                     \binom{n}{j}  
														 d_j - 4(1-r^2)^n\;
				                     \sum_{j=0}^{n} 
				                     \binom{n}{j}  
														 \frac{1}{d_j}
														 \nonumber  \\
														& = & 2^{n+1} - 4(1-r^2)^n\;
				                     \sum_{j=0}^{n} 
				                     \binom{n}{j}  
														 \frac{1}{d_j}.
\end{eqnarray}
It will be shown that  
\begin{equation}
	d_j \geqslant 2(1-r^2)^{n-1}
	\label{eq:dbound}
\end{equation}
and thus
\begin{eqnarray}
  4(1-r^2)^n
	\sum_{j=0}^{n} 
	\binom{n}{j}  
	\frac{1}{d_j} & \leqslant & 
	               2(1-r^2)
	                \sum_{j=0}^{n} 
	                \binom{n}{j}
									\nonumber \\
								& = & 2^{n+1}(1-r^2).
\end{eqnarray}
This will establishes the proposition. 

It remains to prove Eq.~\eqref{eq:dbound}. It is only necessary to show this for $j \geqslant n/2$ since $d_{n-j} = d_j.$ First, for $n=2,$ there are two cases. With $j=1,$ $d_1 = 2 (1-r^2)$  and with $j=2,$ $d_2= 2(1+r^2).$ Eq.~\eqref{eq:dbound} is clearly satisfied for both cases. Suppose that $n> 2$  and let 
\begin{equation}
	\phi(r):= 2 \frac{(1-r^2)^{n-1}}{d_j.}
\end{equation}
Then if $j=n/2,$ $d_j = 2(1-r^2)^{n/2}$ and $\phi(r)  = (1-r^2)^{n/2-1} \leqslant 1$ with equality if and only if $r=0.$ 
Suppose that $j>(n+1)/2.$ Then we shall show that $\phi(r)$ is a monotonically decreasing function of $r$ for $0<r<1.$ First  $\phi(0) = 1$ and $\lim_{r\to 1}\phi(r) = 0.$ Now consider 
\begin{equation}
	\frac{d \phi}{d r} = \frac{(1-r^2)^{n-2}}{d_j^2}\;
	                                   \left[
																		   r(2-n)d_j + (n-2j)c_j
																		 \right].
\end{equation}
This is a continuous function of $r$ and if $\phi(r)$ is not monotonically decreasing or monotonically increasing, then  $\frac{d \phi}{d r}$ must equal zero at some point in the range  $0<r<1.$ Then $\frac{d \phi}{d r} =0$ if and only if
$r(2-n)d_j + (n-2j)c_j.$ This is equivalent to  $r(2-n)\left( 1+ \alpha^{2j-n}\right) = (2j-n)\left( 1- \alpha^{2j-n} \right)$ where $\alpha = (1-r)/(1+r).$ Clearly $\alpha < 1.$ But if $n > 2$ and $j>n/2$ then the left hand side is strictly less than $0$ while the right is positive or $0$. This is impossible and $\frac{\partial \phi}{\partial r} \neq 0.$ Thus $\frac{d \phi}{d r} \neq 0$ and has the same sign throughout the range $0<r<1.$ So $\phi(r)$ is either monotonically decreasing or monotonically increasing. The latter is true since $\phi(1) < \phi(0).$ This shows that $\phi(r) < \phi(0) = 1$ and proves Eq.~\eqref{eq:dbound}.
\hspace{\fill}$\Box$

The results of proposition~4 together with Eq.~\eqref{eq:gengainapprox} imply that for $0<r<1,$
\begin{equation}
	G(\lambda) > m(1-2\lambda)^{2m-2}
	                     \frac{1 - (1-2\lambda)^2 r^2}{1 - (1-2\lambda)^{2m} r^2}.
\end{equation}

It follows that with a single invocation of the phase-flip operation (i.e.\ $m=1$), $G(\lambda) > 1$ whenever $0<r<1.$



\section{Properties of ${\cal Q}$}
\label{app:discordprop}

Straightforward differentiation and algebra shows that 
\begin{equation}
	\frac{\partial{\cal Q}}{\partial \mu} >0
\end{equation}
whenever $\mu>0.$

If $r \geqslant \mu >0$ then 
\begin{eqnarray}
	\frac{\partial{\cal Q}}{\partial r} & = & \frac{r}{2}\; \log_2{\left[ 
	                                                                        \frac{(1+r^2+2r\mu)(1+r^2-2r\mu)}{(1+r^2)^2}
	                                                                      \right]}
																															\nonumber \\
																								 &  & + \frac{\mu}{2}\; \log_2{\left[ 
	                                                                        \frac{(1+r^2+2r\mu)}{(1+r^2-2r\mu)}
	                                                                      \right]}
																															\nonumber \\
																								& \geqslant &  \frac{\mu}{2}\;\log_2{\left[ 
	                                                                        \frac{(1+r^2+2r\mu)(1+r^2-2r\mu)}{(1+r^2)^2}
	                                                                      \right]}
																															\nonumber \\
																								 &  & + \frac{\mu}{2}\; \log_2{\left[ 
	                                                                        \frac{(1+r^2+2r\mu)}{(1+r^2-2r\mu)}
	                                                                      \right]}
																															\nonumber \\
																							  & = & \frac{\mu}{2}\; \log_2{\left[ 
	                                                                        \frac{(1+r^2+2r\mu)^2}{(1+r^2)^2}
	                                                                      \right]}.
\end{eqnarray}
The argument of this logarithm is larger than or equal to $1$ (only possible if $r=0$ or $\mu=0$). This shows that if $r>\mu$ then 	$\frac{\partial{\cal Q}}{\partial r} >0.$				

On the other hand if $\mu >0$ and $r \leqslant \mu$		then 
\begin{eqnarray}
	\frac{\partial{\cal Q}}{\partial r} & = & \frac{r}{2}\; \log_2{\left[ 
	                                                                        \frac{(1+r^2+2r\mu)(1+r^2-2r\mu)}{(1+r^2)^2}
	                                                                      \right]}
																															\nonumber \\
																								 &  & + \frac{\mu}{2}\; \log_2{\left[ 
	                                                                        \frac{(1+r^2+2r\mu)(1-r\mu)}{(1+r^2-2r\mu)(1+r\mu)}
	                                                                      \right]}
																															\nonumber \\
																								& \geqslant &   \frac{r}{2}\; \log_2{\left[ 
	                                                                        \frac{(1+r^2+2r\mu)^2(1-r\mu)}{(1-r^2)^2(1+r\mu)}
	                                                                      \right]}.
\end{eqnarray}
Let 
\begin{equation}
	\omega:= (1+r^2+2r\mu)^2(1-r\mu) - (1-r^2)^2(1+r\mu).
\end{equation}
Then 
\begin{eqnarray}
	\omega & = & 2r\left[ 
	                    \mu(1-r^2)(1+r^2)
											+ 2r(1-r^2\mu^2)
											\right. \nonumber \\
							& & 	\left.
											+ 2r^2\mu(1-mu^2)
	                  \right]
\end{eqnarray}
and for every $0 \leqslant r,mu \leqslant 1$ every term here is positive. Thus $\omega >0$ and the argument of the last logarithm is larger than $1$. Thus if $\mu >0$ and $r \leqslant \mu$	then 	$\frac{\partial{\cal Q}}{\partial r} >0.$	This shows that in all cases 	$\frac{\partial{\cal Q}}{\partial r} >0.$	



\begin{thebibliography}{42}%
\makeatletter
\providecommand \@ifxundefined [1]{%
 \@ifx{#1\undefined}
}%
\providecommand \@ifnum [1]{%
 \ifnum #1\expandafter \@firstoftwo
 \else \expandafter \@secondoftwo
 \fi
}%
\providecommand \@ifx [1]{%
 \ifx #1\expandafter \@firstoftwo
 \else \expandafter \@secondoftwo
 \fi
}%
\providecommand \natexlab [1]{#1}%
\providecommand \enquote  [1]{``#1''}%
\providecommand \bibnamefont  [1]{#1}%
\providecommand \bibfnamefont [1]{#1}%
\providecommand \citenamefont [1]{#1}%
\providecommand \href@noop [0]{\@secondoftwo}%
\providecommand \href [0]{\begingroup \@sanitize@url \@href}%
\providecommand \@href[1]{\@@startlink{#1}\@@href}%
\providecommand \@@href[1]{\endgroup#1\@@endlink}%
\providecommand \@sanitize@url [0]{\catcode `\\12\catcode `\$12\catcode
  `\&12\catcode `\#12\catcode `\^12\catcode `\_12\catcode `\%12\relax}%
\providecommand \@@startlink[1]{}%
\providecommand \@@endlink[0]{}%
\providecommand \url  [0]{\begingroup\@sanitize@url \@url }%
\providecommand \@url [1]{\endgroup\@href {#1}{\urlprefix }}%
\providecommand \urlprefix  [0]{URL }%
\providecommand \Eprint [0]{\href }%
\providecommand \doibase [0]{http://dx.doi.org/}%
\providecommand \selectlanguage [0]{\@gobble}%
\providecommand \bibinfo  [0]{\@secondoftwo}%
\providecommand \bibfield  [0]{\@secondoftwo}%
\providecommand \translation [1]{[#1]}%
\providecommand \BibitemOpen [0]{}%
\providecommand \bibitemStop [0]{}%
\providecommand \bibitemNoStop [0]{.\EOS\space}%
\providecommand \EOS [0]{\spacefactor3000\relax}%
\providecommand \BibitemShut  [1]{\csname bibitem#1\endcsname}%
\let\auto@bib@innerbib\@empty
\bibitem [{\citenamefont {Helstrom}(1976)}]{hellstrom76}%
  \BibitemOpen
  \bibfield  {author} {\bibinfo {author} {\bibfnamefont {C.~W.}\ \bibnamefont
  {Helstrom}},\ }\href@noop {} {\emph {\bibinfo {title} {Quantum Detection and
  Estimation Theory}}}\ (\bibinfo  {publisher} {Academic Press},\ \bibinfo
  {address} {New York},\ \bibinfo {year} {1976})\BibitemShut {NoStop}%
\bibitem [{\citenamefont {Caves}\ \emph {et~al.}(1980)\citenamefont {Caves},
  \citenamefont {Thorne}, \citenamefont {Drever}, \citenamefont {Sandberg},\
  and\ \citenamefont {Zimmermann}}]{caves80}%
  \BibitemOpen
  \bibfield  {author} {\bibinfo {author} {\bibfnamefont {C.~M.}\ \bibnamefont
  {Caves}}, \bibinfo {author} {\bibfnamefont {K.~S.}\ \bibnamefont {Thorne}},
  \bibinfo {author} {\bibfnamefont {R.~W.~P.}\ \bibnamefont {Drever}}, \bibinfo
  {author} {\bibfnamefont {V.~D.}\ \bibnamefont {Sandberg}}, \ and\ \bibinfo
  {author} {\bibfnamefont {M.}~\bibnamefont {Zimmermann}},\ }\href {\doibase
  10.1103/RevModPhys.52.341} {\bibfield  {journal} {\bibinfo  {journal} {Rev.
  Mod. Phys.}\ }\textbf {\bibinfo {volume} {52}},\ \bibinfo {pages} {341}
  (\bibinfo {year} {1980})}\BibitemShut {NoStop}%
\bibitem [{\citenamefont {Shapiro}\ and\ \citenamefont
  {Shepard}(1991)}]{shapiro91}%
  \BibitemOpen
  \bibfield  {author} {\bibinfo {author} {\bibfnamefont {J.~H.}\ \bibnamefont
  {Shapiro}}\ and\ \bibinfo {author} {\bibfnamefont {S.~R.}\ \bibnamefont
  {Shepard}},\ }\href {\doibase 10.1103/PhysRevA.43.3795} {\bibfield  {journal}
  {\bibinfo  {journal} {Phys. Rev. A}\ }\textbf {\bibinfo {volume} {43}},\
  \bibinfo {pages} {3795} (\bibinfo {year} {1991})}\BibitemShut {NoStop}%
\bibitem [{\citenamefont {Caves}(1981)}]{caves93}%
  \BibitemOpen
  \bibfield  {author} {\bibinfo {author} {\bibfnamefont {C.~M.}\ \bibnamefont
  {Caves}},\ }\href {\doibase 10.1103/PhysRevD.23.1693} {\bibfield  {journal}
  {\bibinfo  {journal} {Phys. Rev. D}\ }\textbf {\bibinfo {volume} {23}},\
  \bibinfo {pages} {1693} (\bibinfo {year} {1981})}\BibitemShut {NoStop}%
\bibitem [{\citenamefont {Braunstein}\ and\ \citenamefont
  {Caves}(1994)}]{braunstein94}%
  \BibitemOpen
  \bibfield  {author} {\bibinfo {author} {\bibfnamefont {S.~L.}\ \bibnamefont
  {Braunstein}}\ and\ \bibinfo {author} {\bibfnamefont {C.~M.}\ \bibnamefont
  {Caves}},\ }\href {\doibase 10.1103/PhysRevLett.72.3439} {\bibfield
  {journal} {\bibinfo  {journal} {Phys. Rev. Lett.}\ }\textbf {\bibinfo
  {volume} {72}},\ \bibinfo {pages} {3439} (\bibinfo {year}
  {1994})}\BibitemShut {NoStop}%
\bibitem [{\citenamefont {Giovannetti}\ \emph {et~al.}(2004)\citenamefont
  {Giovannetti}, \citenamefont {Lloyd},\ and\ \citenamefont
  {Maccone}}]{giovannetti04}%
  \BibitemOpen
  \bibfield  {author} {\bibinfo {author} {\bibfnamefont {V.}~\bibnamefont
  {Giovannetti}}, \bibinfo {author} {\bibfnamefont {S.}~\bibnamefont {Lloyd}},
  \ and\ \bibinfo {author} {\bibfnamefont {L.}~\bibnamefont {Maccone}},\ }\href
  {\doibase 10.1126/science.1104149} {\bibfield  {journal} {\bibinfo  {journal}
  {Science}\ }\textbf {\bibinfo {volume} {306}},\ \bibinfo {pages} {1330}
  (\bibinfo {year} {2004})}\BibitemShut {NoStop}%
\bibitem [{\citenamefont {Sarovar}\ and\ \citenamefont
  {Milburn}(2006)}]{sarovar06}%
  \BibitemOpen
  \bibfield  {author} {\bibinfo {author} {\bibfnamefont {M.}~\bibnamefont
  {Sarovar}}\ and\ \bibinfo {author} {\bibfnamefont {G.~J.}\ \bibnamefont
  {Milburn}},\ }\href {http://stacks.iop.org/0305-4470/39/i=26/a=015}
  {\bibfield  {journal} {\bibinfo  {journal} {J. Phys. A}\ }\textbf {\bibinfo
  {volume} {39}},\ \bibinfo {pages} {8487} (\bibinfo {year}
  {2006})}\BibitemShut {NoStop}%
\bibitem [{\citenamefont {Giovannetti}\ \emph {et~al.}(2006)\citenamefont
  {Giovannetti}, \citenamefont {Lloyd},\ and\ \citenamefont
  {Maccone}}]{giovannetti06}%
  \BibitemOpen
  \bibfield  {author} {\bibinfo {author} {\bibfnamefont {V.}~\bibnamefont
  {Giovannetti}}, \bibinfo {author} {\bibfnamefont {S.}~\bibnamefont {Lloyd}},
  \ and\ \bibinfo {author} {\bibfnamefont {L.}~\bibnamefont {Maccone}},\ }\href
  {\doibase 10.1103/PhysRevLett.96.010401} {\bibfield  {journal} {\bibinfo
  {journal} {Phys. Rev. Lett.}\ }\textbf {\bibinfo {volume} {96}},\ \bibinfo
  {pages} {010401} (\bibinfo {year} {2006})}\BibitemShut {NoStop}%
\bibitem [{\citenamefont {Paris}(2009)}]{paris09}%
  \BibitemOpen
  \bibfield  {author} {\bibinfo {author} {\bibfnamefont {M.}~\bibnamefont
  {Paris}},\ }\href {\doibase 10.1142/S0219749909004839} {\bibfield  {journal}
  {\bibinfo  {journal} {Int. J. Quantum Information}\ }\textbf {\bibinfo
  {volume} {9}},\ \bibinfo {pages} {125} (\bibinfo {year} {2009})}\BibitemShut
  {NoStop}%
\bibitem [{\citenamefont {Berry}\ \emph {et~al.}(2009)\citenamefont {Berry},
  \citenamefont {Higgins}, \citenamefont {Bartlett}, \citenamefont {Mitchell},
  \citenamefont {Pryde},\ and\ \citenamefont {Wiseman}}]{berry09}%
  \BibitemOpen
  \bibfield  {author} {\bibinfo {author} {\bibfnamefont {D.~W.}\ \bibnamefont
  {Berry}}, \bibinfo {author} {\bibfnamefont {B.~L.}\ \bibnamefont {Higgins}},
  \bibinfo {author} {\bibfnamefont {S.~D.}\ \bibnamefont {Bartlett}}, \bibinfo
  {author} {\bibfnamefont {M.~W.}\ \bibnamefont {Mitchell}}, \bibinfo {author}
  {\bibfnamefont {G.~J.}\ \bibnamefont {Pryde}}, \ and\ \bibinfo {author}
  {\bibfnamefont {H.~M.}\ \bibnamefont {Wiseman}},\ }\href {\doibase
  10.1103/PhysRevA.80.052114} {\bibfield  {journal} {\bibinfo  {journal} {Phys.
  Rev. A}\ }\textbf {\bibinfo {volume} {80}},\ \bibinfo {pages} {052114}
  (\bibinfo {year} {2009})}\BibitemShut {NoStop}%
\bibitem [{\citenamefont {O'Loan}()}]{oloan10}%
  \BibitemOpen
  \bibfield  {author} {\bibinfo {author} {\bibfnamefont {C.}~\bibnamefont
  {O'Loan}},\ }\href@noop {} {\bibfield  {journal} {\bibinfo  {journal}
  {e-print}\ }}\Eprint {http://arxiv.org/abs/arXiv:1001.3971v1}
  {arXiv:1001.3971v1 [quant-th]} \BibitemShut {NoStop}%
\bibitem [{\citenamefont {Escher}\ \emph {et~al.}(2011)\citenamefont {Escher},
  \citenamefont {de~Matos~Filho},\ and\ \citenamefont {Davidovich}}]{escher11}%
  \BibitemOpen
  \bibfield  {author} {\bibinfo {author} {\bibfnamefont {B.~M.}\ \bibnamefont
  {Escher}}, \bibinfo {author} {\bibfnamefont {R.~L.}\ \bibnamefont
  {de~Matos~Filho}}, \ and\ \bibinfo {author} {\bibfnamefont {L.}~\bibnamefont
  {Davidovich}},\ }\href {\doibase 10.1038/nphys1958} {\bibfield  {journal}
  {\bibinfo  {journal} {Nature Phys.}\ }\textbf {\bibinfo {volume} {7}},\
  \bibinfo {pages} {406} (\bibinfo {year} {2011})}\BibitemShut {NoStop}%
\bibitem [{\citenamefont {Giovannetti}\ \emph {et~al.}(2011)\citenamefont
  {Giovannetti}, \citenamefont {Lloyd},\ and\ \citenamefont
  {Maccone}}]{giovannetti11}%
  \BibitemOpen
  \bibfield  {author} {\bibinfo {author} {\bibfnamefont {V.}~\bibnamefont
  {Giovannetti}}, \bibinfo {author} {\bibfnamefont {S.}~\bibnamefont {Lloyd}},
  \ and\ \bibinfo {author} {\bibfnamefont {L.}~\bibnamefont {Maccone}},\ }\href
  {\doibase doi:10.1038/nphoton.2011.35} {\bibfield  {journal} {\bibinfo
  {journal} {Nature Photonics}\ }\textbf {\bibinfo {volume} {5}},\ \bibinfo
  {pages} {222} (\bibinfo {year} {2011})}\BibitemShut {NoStop}%
\bibitem [{\citenamefont {Nielsen}\ and\ \citenamefont
  {Chuang}(2000)}]{nielsen00}%
  \BibitemOpen
  \bibfield  {author} {\bibinfo {author} {\bibfnamefont {M.~A.}\ \bibnamefont
  {Nielsen}}\ and\ \bibinfo {author} {\bibfnamefont {I.~L.}\ \bibnamefont
  {Chuang}},\ }\href@noop {} {\emph {\bibinfo {title} {Quantum Computation and
  Quantum Information}}}\ (\bibinfo  {publisher} {Cambridge University Press},\
  \bibinfo {address} {Cambridge},\ \bibinfo {year} {2000})\BibitemShut
  {NoStop}%
\bibitem [{\citenamefont {Slichter}(1996)}]{slichter96}%
  \BibitemOpen
  \bibfield  {author} {\bibinfo {author} {\bibfnamefont {C.~P.}\ \bibnamefont
  {Slichter}},\ }\href@noop {} {\emph {\bibinfo {title} {Principles of Magnetic
  Resonance}}}\ (\bibinfo  {publisher} {Springer},\ \bibinfo {address}
  {Berlin},\ \bibinfo {year} {1996})\BibitemShut {NoStop}%
\bibitem [{\citenamefont {Jones}(2011)}]{jones11}%
  \BibitemOpen
  \bibfield  {author} {\bibinfo {author} {\bibfnamefont {J.~A.}\ \bibnamefont
  {Jones}},\ }\href {\doibase http://dx.doi.org/10.1016/j.pnmrs.2010.11.001}
  {\bibfield  {journal} {\bibinfo  {journal} {Prog. NMR Spectrosc.}\ }\textbf
  {\bibinfo {volume} {59}},\ \bibinfo {pages} {91} (\bibinfo {year}
  {2011})}\BibitemShut {NoStop}%
\bibitem [{\citenamefont {Zwierz}\ \emph {et~al.}(2010)\citenamefont {Zwierz},
  \citenamefont {P\'erez-Delgado},\ and\ \citenamefont {Kok}}]{zwierz10}%
  \BibitemOpen
  \bibfield  {author} {\bibinfo {author} {\bibfnamefont {M.}~\bibnamefont
  {Zwierz}}, \bibinfo {author} {\bibfnamefont {C.~A.}\ \bibnamefont
  {P\'erez-Delgado}}, \ and\ \bibinfo {author} {\bibfnamefont {P.}~\bibnamefont
  {Kok}},\ }\href {\doibase 10.1103/PhysRevLett.105.180402} {\bibfield
  {journal} {\bibinfo  {journal} {Phys. Rev. Lett.}\ }\textbf {\bibinfo
  {volume} {105}},\ \bibinfo {pages} {180402} (\bibinfo {year}
  {2010})}\BibitemShut {NoStop}%
\bibitem [{\citenamefont {Zwierz}\ \emph {et~al.}(2012)\citenamefont {Zwierz},
  \citenamefont {P\'erez-Delgado},\ and\ \citenamefont {Kok}}]{zwierz12}%
  \BibitemOpen
  \bibfield  {author} {\bibinfo {author} {\bibfnamefont {M.}~\bibnamefont
  {Zwierz}}, \bibinfo {author} {\bibfnamefont {C.~A.}\ \bibnamefont
  {P\'erez-Delgado}}, \ and\ \bibinfo {author} {\bibfnamefont {P.}~\bibnamefont
  {Kok}},\ }\href {\doibase 10.1103/PhysRevA.85.042112} {\bibfield  {journal}
  {\bibinfo  {journal} {Phys. Rev. A}\ }\textbf {\bibinfo {volume} {85}},\
  \bibinfo {pages} {042112} (\bibinfo {year} {2012})}\BibitemShut {NoStop}%
\bibitem [{\citenamefont {Fujiwara}\ and\ \citenamefont
  {Imai}(2003)}]{fujiwara03}%
  \BibitemOpen
  \bibfield  {author} {\bibinfo {author} {\bibfnamefont {A.}~\bibnamefont
  {Fujiwara}}\ and\ \bibinfo {author} {\bibfnamefont {H.}~\bibnamefont
  {Imai}},\ }\href {http://stacks.iop.org/0305-4470/36/i=29/a=314} {\bibfield
  {journal} {\bibinfo  {journal} {J. Phys. A}\ }\textbf {\bibinfo {volume}
  {36}},\ \bibinfo {pages} {8093} (\bibinfo {year} {2003})}\BibitemShut
  {NoStop}%
\bibitem [{\citenamefont {O'Loan}(2007)}]{oloan07}%
  \BibitemOpen
  \bibfield  {author} {\bibinfo {author} {\bibfnamefont {C.~J.}\ \bibnamefont
  {O'Loan}},\ }\href {http://stacks.iop.org/1751-8121/40/i=48/a=013} {\bibfield
   {journal} {\bibinfo  {journal} {J. Phys. A}\ }\textbf {\bibinfo {volume}
  {40}},\ \bibinfo {pages} {14499} (\bibinfo {year} {2007})}\BibitemShut
  {NoStop}%
\bibitem [{\citenamefont {Ji}\ \emph {et~al.}(2008)\citenamefont {Ji},
  \citenamefont {Wang}, \citenamefont {Duan}, \citenamefont {Feng},\ and\
  \citenamefont {Ying}}]{ji08}%
  \BibitemOpen
  \bibfield  {author} {\bibinfo {author} {\bibfnamefont {Z.}~\bibnamefont
  {Ji}}, \bibinfo {author} {\bibfnamefont {G.}~\bibnamefont {Wang}}, \bibinfo
  {author} {\bibfnamefont {R.}~\bibnamefont {Duan}}, \bibinfo {author}
  {\bibfnamefont {Y.}~\bibnamefont {Feng}}, \ and\ \bibinfo {author}
  {\bibfnamefont {M.}~\bibnamefont {Ying}},\ }\href {\doibase
  http://dx.doi.org/10.1109/TIT.2008.929940} {\bibfield  {journal} {\bibinfo
  {journal} {IEEE Trans. Inf. Theory}\ }\textbf {\bibinfo {volume} {54}},\
  \bibinfo {pages} {5172 } (\bibinfo {year} {2008})}\BibitemShut {NoStop}%
\bibitem [{\citenamefont {Chiuri}\ \emph {et~al.}(2011)\citenamefont {Chiuri},
  \citenamefont {Rosati}, \citenamefont {Vallone}, \citenamefont {P\'adua},
  \citenamefont {Imai}, \citenamefont {Giacomini}, \citenamefont
  {Macchiavello},\ and\ \citenamefont {Mataloni}}]{chiuri11}%
  \BibitemOpen
  \bibfield  {author} {\bibinfo {author} {\bibfnamefont {A.}~\bibnamefont
  {Chiuri}}, \bibinfo {author} {\bibfnamefont {V.}~\bibnamefont {Rosati}},
  \bibinfo {author} {\bibfnamefont {G.}~\bibnamefont {Vallone}}, \bibinfo
  {author} {\bibfnamefont {S.}~\bibnamefont {P\'adua}}, \bibinfo {author}
  {\bibfnamefont {H.}~\bibnamefont {Imai}}, \bibinfo {author} {\bibfnamefont
  {S.}~\bibnamefont {Giacomini}}, \bibinfo {author} {\bibfnamefont
  {C.}~\bibnamefont {Macchiavello}}, \ and\ \bibinfo {author} {\bibfnamefont
  {P.}~\bibnamefont {Mataloni}},\ }\href {\doibase
  10.1103/PhysRevLett.107.253602} {\bibfield  {journal} {\bibinfo  {journal}
  {Phys. Rev. Lett.}\ }\textbf {\bibinfo {volume} {107}},\ \bibinfo {pages}
  {253602} (\bibinfo {year} {2011})}\BibitemShut {NoStop}%
\bibitem [{\citenamefont {Ruppert}\ \emph {et~al.}(2012)\citenamefont
  {Ruppert}, \citenamefont {Virosztek},\ and\ \citenamefont
  {Hangos}}]{ruppert12}%
  \BibitemOpen
  \bibfield  {author} {\bibinfo {author} {\bibfnamefont {L.}~\bibnamefont
  {Ruppert}}, \bibinfo {author} {\bibfnamefont {D.}~\bibnamefont {Virosztek}},
  \ and\ \bibinfo {author} {\bibfnamefont {K.}~\bibnamefont {Hangos}},\ }\href
  {http://stacks.iop.org/1751-8121/45/i=26/a=265305} {\bibfield  {journal}
  {\bibinfo  {journal} {J Phys. A}\ }\textbf {\bibinfo {volume} {45}},\
  \bibinfo {pages} {265305} (\bibinfo {year} {2012})}\BibitemShut {NoStop}%
\bibitem [{\citenamefont {D'Ariano}\ \emph {et~al.}(2005)\citenamefont
  {D'Ariano}, \citenamefont {Macchiavello},\ and\ \citenamefont
  {Perinotti}}]{dariano05}%
  \BibitemOpen
  \bibfield  {author} {\bibinfo {author} {\bibfnamefont {G.~M.}\ \bibnamefont
  {D'Ariano}}, \bibinfo {author} {\bibfnamefont {C.}~\bibnamefont
  {Macchiavello}}, \ and\ \bibinfo {author} {\bibfnamefont {P.}~\bibnamefont
  {Perinotti}},\ }\href {\doibase 10.1103/PhysRevA.72.042327} {\bibfield
  {journal} {\bibinfo  {journal} {Phys. Rev. A}\ }\textbf {\bibinfo {volume}
  {72}},\ \bibinfo {pages} {042327} (\bibinfo {year} {2005})}\BibitemShut
  {NoStop}%
\bibitem [{\citenamefont {Boixo}\ and\ \citenamefont {Somma}(2008)}]{boixo08b}%
  \BibitemOpen
  \bibfield  {author} {\bibinfo {author} {\bibfnamefont {S.}~\bibnamefont
  {Boixo}}\ and\ \bibinfo {author} {\bibfnamefont {R.~D.}\ \bibnamefont
  {Somma}},\ }\href {\doibase 10.1103/PhysRevA.77.052320} {\bibfield  {journal}
  {\bibinfo  {journal} {Phys. Rev. A}\ }\textbf {\bibinfo {volume} {77}},\
  \bibinfo {pages} {052320} (\bibinfo {year} {2008})}\BibitemShut {NoStop}%
\bibitem [{\citenamefont {Modi}\ \emph {et~al.}(2011)\citenamefont {Modi},
  \citenamefont {Cable}, \citenamefont {Williamson},\ and\ \citenamefont
  {Vedral}}]{modi11}%
  \BibitemOpen
  \bibfield  {author} {\bibinfo {author} {\bibfnamefont {K.}~\bibnamefont
  {Modi}}, \bibinfo {author} {\bibfnamefont {H.}~\bibnamefont {Cable}},
  \bibinfo {author} {\bibfnamefont {M.}~\bibnamefont {Williamson}}, \ and\
  \bibinfo {author} {\bibfnamefont {V.}~\bibnamefont {Vedral}},\ }\href
  {\doibase 10.1103/PhysRevX.1.021022} {\bibfield  {journal} {\bibinfo
  {journal} {Phys. Rev. X}\ }\textbf {\bibinfo {volume} {1}},\ \bibinfo {pages}
  {021022} (\bibinfo {year} {2011})}\BibitemShut {NoStop}%
\bibitem [{\citenamefont {Datta}\ \emph {et~al.}(2011)\citenamefont {Datta},
  \citenamefont {Zhang}, \citenamefont {Thomas-Peter}, \citenamefont {Dorner},
  \citenamefont {Smith},\ and\ \citenamefont {Walmsley}}]{datta11}%
  \BibitemOpen
  \bibfield  {author} {\bibinfo {author} {\bibfnamefont {A.}~\bibnamefont
  {Datta}}, \bibinfo {author} {\bibfnamefont {L.}~\bibnamefont {Zhang}},
  \bibinfo {author} {\bibfnamefont {N.}~\bibnamefont {Thomas-Peter}}, \bibinfo
  {author} {\bibfnamefont {U.}~\bibnamefont {Dorner}}, \bibinfo {author}
  {\bibfnamefont {B.~J.}\ \bibnamefont {Smith}}, \ and\ \bibinfo {author}
  {\bibfnamefont {I.~A.}\ \bibnamefont {Walmsley}},\ }\href {\doibase
  10.1103/PhysRevA.83.063836} {\bibfield  {journal} {\bibinfo  {journal} {Phys.
  Rev. A}\ }\textbf {\bibinfo {volume} {83}},\ \bibinfo {pages} {063836}
  (\bibinfo {year} {2011})}\BibitemShut {NoStop}%
\bibitem [{\citenamefont {van Dam}\ \emph {et~al.}(2007)\citenamefont {van
  Dam}, \citenamefont {D'Ariano}, \citenamefont {Ekert}, \citenamefont
  {Macchiavello},\ and\ \citenamefont {Mosca}}]{vandam07}%
  \BibitemOpen
  \bibfield  {author} {\bibinfo {author} {\bibfnamefont {W.}~\bibnamefont {van
  Dam}}, \bibinfo {author} {\bibfnamefont {G.~M.}\ \bibnamefont {D'Ariano}},
  \bibinfo {author} {\bibfnamefont {A.}~\bibnamefont {Ekert}}, \bibinfo
  {author} {\bibfnamefont {C.}~\bibnamefont {Macchiavello}}, \ and\ \bibinfo
  {author} {\bibfnamefont {M.}~\bibnamefont {Mosca}},\ }\href {\doibase
  10.1103/PhysRevLett.98.090501} {\bibfield  {journal} {\bibinfo  {journal}
  {Phys. Rev. Lett.}\ }\textbf {\bibinfo {volume} {98}},\ \bibinfo {pages}
  {090501} (\bibinfo {year} {2007})}\BibitemShut {NoStop}%
\bibitem [{\citenamefont {Cram\'{e}r}(1946)}]{cramer46}%
  \BibitemOpen
  \bibfield  {author} {\bibinfo {author} {\bibfnamefont {H.}~\bibnamefont
  {Cram\'{e}r}},\ }\href@noop {} {\emph {\bibinfo {title} {Mathematical Methods
  of Statistics}}}\ (\bibinfo  {publisher} {Princeton University Press},\
  \bibinfo {address} {Princeton, NJ},\ \bibinfo {year} {1946})\BibitemShut
  {NoStop}%
\bibitem [{\citenamefont {Barndorff-Nielsen}\ and\ \citenamefont
  {Gill}(2000)}]{barndorff00}%
  \BibitemOpen
  \bibfield  {author} {\bibinfo {author} {\bibfnamefont {O.~E.}\ \bibnamefont
  {Barndorff-Nielsen}}\ and\ \bibinfo {author} {\bibfnamefont {R.~D.}\
  \bibnamefont {Gill}},\ }\href {http://stacks.iop.org/0305-4470/33/i=24/a=306}
  {\bibfield  {journal} {\bibinfo  {journal} {J. Phys A: Math. Gen.}\ }\textbf
  {\bibinfo {volume} {33}},\ \bibinfo {pages} {4481} (\bibinfo {year}
  {2000})}\BibitemShut {NoStop}%
\bibitem [{\citenamefont {Collins}\ \emph {et~al.}(2000)\citenamefont
  {Collins}, \citenamefont {Kim}, \citenamefont {Holton}, \citenamefont
  {Sierzputowska-Gracz},\ and\ \citenamefont {Stejskal}}]{collins00}%
  \BibitemOpen
  \bibfield  {author} {\bibinfo {author} {\bibfnamefont {D.}~\bibnamefont
  {Collins}}, \bibinfo {author} {\bibfnamefont {K.~W.}\ \bibnamefont {Kim}},
  \bibinfo {author} {\bibfnamefont {W.~C.}\ \bibnamefont {Holton}}, \bibinfo
  {author} {\bibfnamefont {H.}~\bibnamefont {Sierzputowska-Gracz}}, \ and\
  \bibinfo {author} {\bibfnamefont {E.~O.}\ \bibnamefont {Stejskal}},\ }\href
  {\doibase 10.1103/PhysRevA.62.022304} {\bibfield  {journal} {\bibinfo
  {journal} {Phys. Rev. A}\ }\textbf {\bibinfo {volume} {62}},\ \bibinfo
  {pages} {022304} (\bibinfo {year} {2000})}\BibitemShut {NoStop}%
\bibitem [{\citenamefont {Peres}(1996)}]{peres96}%
  \BibitemOpen
  \bibfield  {author} {\bibinfo {author} {\bibfnamefont {A.}~\bibnamefont
  {Peres}},\ }\href {\doibase 10.1103/PhysRevLett.77.1413} {\bibfield
  {journal} {\bibinfo  {journal} {Phys. Rev. Lett.}\ }\textbf {\bibinfo
  {volume} {77}},\ \bibinfo {pages} {1413} (\bibinfo {year}
  {1996})}\BibitemShut {NoStop}%
\bibitem [{\citenamefont {Horodecki}\ \emph {et~al.}(1996)\citenamefont
  {Horodecki}, \citenamefont {Horodecki},\ and\ \citenamefont
  {Horodecki}}]{horodecki96}%
  \BibitemOpen
  \bibfield  {author} {\bibinfo {author} {\bibfnamefont {M.}~\bibnamefont
  {Horodecki}}, \bibinfo {author} {\bibfnamefont {P.}~\bibnamefont
  {Horodecki}}, \ and\ \bibinfo {author} {\bibfnamefont {R.}~\bibnamefont
  {Horodecki}},\ }\href {\doibase 10.1016/S0375-9601(96)00706-2} {\bibfield
  {journal} {\bibinfo  {journal} {Phys. Lett. A}\ }\textbf {\bibinfo {volume}
  {223}},\ \bibinfo {pages} {1 } (\bibinfo {year} {1996})}\BibitemShut
  {NoStop}%
\bibitem [{\citenamefont {Ollivier}\ and\ \citenamefont
  {Zurek}(2001)}]{ollivier01}%
  \BibitemOpen
  \bibfield  {author} {\bibinfo {author} {\bibfnamefont {H.}~\bibnamefont
  {Ollivier}}\ and\ \bibinfo {author} {\bibfnamefont {W.~H.}\ \bibnamefont
  {Zurek}},\ }\href {\doibase 10.1103/PhysRevLett.88.017901} {\bibfield
  {journal} {\bibinfo  {journal} {Phys. Rev. Lett.}\ }\textbf {\bibinfo
  {volume} {88}},\ \bibinfo {pages} {017901} (\bibinfo {year}
  {2001})}\BibitemShut {NoStop}%
\bibitem [{\citenamefont {Modi}\ \emph {et~al.}()\citenamefont {Modi},
  \citenamefont {Brodutch}, \citenamefont {Cable}, \citenamefont {Paterek},\
  and\ \citenamefont {Vedral}}]{modi11a}%
  \BibitemOpen
  \bibfield  {author} {\bibinfo {author} {\bibfnamefont {K.}~\bibnamefont
  {Modi}}, \bibinfo {author} {\bibfnamefont {A.}~\bibnamefont {Brodutch}},
  \bibinfo {author} {\bibfnamefont {H.}~\bibnamefont {Cable}}, \bibinfo
  {author} {\bibfnamefont {T.}~\bibnamefont {Paterek}}, \ and\ \bibinfo
  {author} {\bibfnamefont {V.}~\bibnamefont {Vedral}},\ }\href@noop {}
  {\bibfield  {journal} {\bibinfo  {journal} {e-print}\ }}\Eprint
  {http://arxiv.org/abs/arXiv:1112.6238} {arXiv:1112.6238 [quant-th]}
  \BibitemShut {NoStop}%
\bibitem [{\citenamefont {Luo}(2008)}]{luo08}%
  \BibitemOpen
  \bibfield  {author} {\bibinfo {author} {\bibfnamefont {S.}~\bibnamefont
  {Luo}},\ }\href {\doibase 10.1103/PhysRevA.77.042303} {\bibfield  {journal}
  {\bibinfo  {journal} {Phys. Rev. A}\ }\textbf {\bibinfo {volume} {77}},\
  \bibinfo {pages} {042303} (\bibinfo {year} {2008})}\BibitemShut {NoStop}%
\bibitem [{\citenamefont {Chen}\ \emph {et~al.}(2011)\citenamefont {Chen},
  \citenamefont {Zhang}, \citenamefont {Yu}, \citenamefont {Yi},\ and\
  \citenamefont {Oh}}]{chen11}%
  \BibitemOpen
  \bibfield  {author} {\bibinfo {author} {\bibfnamefont {Q.}~\bibnamefont
  {Chen}}, \bibinfo {author} {\bibfnamefont {C.}~\bibnamefont {Zhang}},
  \bibinfo {author} {\bibfnamefont {S.}~\bibnamefont {Yu}}, \bibinfo {author}
  {\bibfnamefont {X.~X.}\ \bibnamefont {Yi}}, \ and\ \bibinfo {author}
  {\bibfnamefont {C.~H.}\ \bibnamefont {Oh}},\ }\href {\doibase
  10.1103/PhysRevA.84.042313} {\bibfield  {journal} {\bibinfo  {journal} {Phys.
  Rev. A}\ }\textbf {\bibinfo {volume} {84}},\ \bibinfo {pages} {042313}
  (\bibinfo {year} {2011})}\BibitemShut {NoStop}%
\bibitem [{\citenamefont {Marx}\ \emph {et~al.}(2000)\citenamefont {Marx},
  \citenamefont {Fahmy}, \citenamefont {Myers}, \citenamefont {Bermel},\ and\
  \citenamefont {Glaser}}]{marx00}%
  \BibitemOpen
  \bibfield  {author} {\bibinfo {author} {\bibfnamefont {R.}~\bibnamefont
  {Marx}}, \bibinfo {author} {\bibfnamefont {A.~F.}\ \bibnamefont {Fahmy}},
  \bibinfo {author} {\bibfnamefont {J.~M.}\ \bibnamefont {Myers}}, \bibinfo
  {author} {\bibfnamefont {W.}~\bibnamefont {Bermel}}, \ and\ \bibinfo {author}
  {\bibfnamefont {S.~J.}\ \bibnamefont {Glaser}},\ }\href {\doibase
  10.1103/PhysRevA.62.012310} {\bibfield  {journal} {\bibinfo  {journal} {Phys.
  Rev. A}\ }\textbf {\bibinfo {volume} {62}},\ \bibinfo {pages} {012310}
  (\bibinfo {year} {2000})}\BibitemShut {NoStop}%
\bibitem [{\citenamefont {Das}\ \emph {et~al.}(2002)\citenamefont {Das},
  \citenamefont {Mahesh},\ and\ \citenamefont {Kumar}}]{das02}%
  \BibitemOpen
  \bibfield  {author} {\bibinfo {author} {\bibfnamefont {R.}~\bibnamefont
  {Das}}, \bibinfo {author} {\bibfnamefont {T.}~\bibnamefont {Mahesh}}, \ and\
  \bibinfo {author} {\bibfnamefont {A.}~\bibnamefont {Kumar}},\ }\href
  {\doibase http://dx.doi.org/10.1016/S1090-7807(02)00009-5} {\bibfield
  {journal} {\bibinfo  {journal} {J. Mag. Res.}\ }\textbf {\bibinfo {volume}
  {159}},\ \bibinfo {pages} {46} (\bibinfo {year} {2002})}\BibitemShut
  {NoStop}%
\bibitem [{\citenamefont {Peng}\ \emph {et~al.}(2002)\citenamefont {Peng},
  \citenamefont {Zhu}, \citenamefont {Fang}, \citenamefont {Feng},
  \citenamefont {Liu},\ and\ \citenamefont {Gao}}]{peng02}%
  \BibitemOpen
  \bibfield  {author} {\bibinfo {author} {\bibfnamefont {X.}~\bibnamefont
  {Peng}}, \bibinfo {author} {\bibfnamefont {X.}~\bibnamefont {Zhu}}, \bibinfo
  {author} {\bibfnamefont {X.}~\bibnamefont {Fang}}, \bibinfo {author}
  {\bibfnamefont {M.}~\bibnamefont {Feng}}, \bibinfo {author} {\bibfnamefont
  {M.}~\bibnamefont {Liu}}, \ and\ \bibinfo {author} {\bibfnamefont
  {K.}~\bibnamefont {Gao}},\ }\href {\doibase 10.1103/PhysRevA.65.042315}
  {\bibfield  {journal} {\bibinfo  {journal} {Phys. Rev. A}\ }\textbf {\bibinfo
  {volume} {65}},\ \bibinfo {pages} {042315} (\bibinfo {year}
  {2002})}\BibitemShut {NoStop}%
\bibitem [{\citenamefont {Ramanathan}\ \emph {et~al.}(2004)\citenamefont
  {Ramanathan}, \citenamefont {Boulant}, \citenamefont {Chen}, \citenamefont
  {Cory}, \citenamefont {Chuang},\ and\ \citenamefont
  {Steffen}}]{ramanathan04}%
  \BibitemOpen
  \bibfield  {author} {\bibinfo {author} {\bibfnamefont {C.}~\bibnamefont
  {Ramanathan}}, \bibinfo {author} {\bibfnamefont {N.}~\bibnamefont {Boulant}},
  \bibinfo {author} {\bibfnamefont {Z.}~\bibnamefont {Chen}}, \bibinfo {author}
  {\bibfnamefont {D.~G.}\ \bibnamefont {Cory}}, \bibinfo {author}
  {\bibfnamefont {I.}~\bibnamefont {Chuang}}, \ and\ \bibinfo {author}
  {\bibfnamefont {M.}~\bibnamefont {Steffen}},\ }\href
  {http://dx.doi.org/10.1007/s11128-004-3668-x} {\bibfield  {journal} {\bibinfo
   {journal} {Quantum Information Processing}\ }\textbf {\bibinfo {volume}
  {3}},\ \bibinfo {pages} {15} (\bibinfo {year} {2004})},\ \bibinfo {note}
  {10.1007/s11128-004-3668-x}\BibitemShut {NoStop}%
\bibitem [{\citenamefont {Xiao}\ and\ \citenamefont {Jones}(2005)}]{xiao06}%
  \BibitemOpen
  \bibfield  {author} {\bibinfo {author} {\bibfnamefont {L.}~\bibnamefont
  {Xiao}}\ and\ \bibinfo {author} {\bibfnamefont {J.~A.}\ \bibnamefont
  {Jones}},\ }\href {\doibase 10.1103/PhysRevA.72.032326} {\bibfield  {journal}
  {\bibinfo  {journal} {Phys. Rev. A}\ }\textbf {\bibinfo {volume} {72}},\
  \bibinfo {eid} {032326} (\bibinfo {year} {2005})}\BibitemShut {NoStop}%
\end{thebibliography}

%

\end{document}